\documentclass[printer]{aa}
\usepackage{natbib}
\usepackage[table]{xcolor}
\usepackage[per-mode=symbol]{siunitx}
 \DeclareSIUnit\angstrom{\text {Å}}
 \DeclareSIUnit\gauss{G}
 \DeclareSIUnit\kilogauss{kG}
 \DeclareSIUnit\maxwell{Mx}
 \DeclareSIUnit\mh{MHS}
\usepackage{amsmath}
\usepackage{nccmath}
\usepackage{graphicx }
\usepackage{soul}

\begin{document}

\title{Decay of a photospheric transient filament at the boundary of a pore and the chromospheric response}

\author{P. Lindner \inst{1} \and R. Schlichenmaier \inst{1} \and N. Bello Gonz\'alez \inst{1} \and J. de la Cruz Rodr{\'\i}guez \inst{2}}

\titlerunning{Evolution of a transient filament adjecent to a pore and a chromospheric response}
\authorrunning{Lindner et al.} 

\institute{
Leibniz-Institut für Sonnenphysik (KIS), Schöneckstr. 6, 79104 Freiburg, Germany 
\and 
Institute for Solar Physics, Dept. of Astronomy, Stockholm University, AlbaNova University Centre, 106 91, Stockholm, Sweden
}

\abstract{Intermediate stages between pores and sunspots are a rare phenomenon and can manifest with the formation of transient photospheric penumbral-like filaments. Although the magnetic field changes rapidly during the evolution of such filaments, they have not been shown to be connected to magnetic reconnection events yet.}
{We characterize the evolution of transient photospheric filaments around a pore and search for possible signs of chromospheric responses.}
{We analyzed observations of a pore in NOAA~AR~12739 from the Swedish Solar Telescope including spectropolarimetric data of the \ion{Fe}{I}~6173~\SI{}{\angstrom} and the \ion{Ca}{II}~8542~\SI{}{\angstrom} line and spectroscopic data of the \ion{Ca}{II}~K~3934~\SI{}{\angstrom} line. The VFISV Milne-Eddington inversion code and the multi-line Non-LTE inversion code STiC were utilized to obtain atmospheric parameters in the photosphere and the chromosphere.}
{Multiple filamentary structures of inclined magnetic fields are found in photospheric inclination maps 
at the boundary of the pore, although the pore never developed a penumbra. One of the filaments shows a clear counterpart in continuum intensity maps in addition to photospheric blueshifts. During its decay, a brightening in the blue wing of the \ion{Ca}{II}~8542~\SI{}{\angstrom} line is observed. The \ion{Ca}{II}~K~3934~\SI{}{\angstrom} and the \ion{Ca}{II}~8542~\SI{}{\angstrom} lines show complex spectral profiles in this region. Depth-dependent STiC inversion results using data from all available lines yield a temperature increase (roughly \SI{1000}{\kelvin}) and bidirectional flows (magnitudes up to \SI{8}{\kilo \meter \per \second}) at $\log\tau=-3.5$.}
{The temporal and spatial correlation of the decaying filament (observed in the photosphere) to the temperature increase and the bidirectional flows in the high photosphere/low chromosphere suggests that they are connected. We propose scenarios in which magnetic reconnection happens at the edge of a rising magnetic flux tube in the photosphere. This leads to both the decay of the filament in the photosphere and the observed temperature increase and the bidirectional flows in the high photosphere/low chromosphere.} 

\keywords{sunspots, Sun: chromosphere, Sun: photosphere,Sun: magnetic fields, Sun: activity}

\maketitle

\section{Introduction}
Pores and sunspots are manifestations of magnetic fields that penetrate the solar surface. The presence of these magnetic fields inhibits the granular convection which would otherwise take place as in the undisturbed quiet solar regions. Since less hot material is transported from deeper layers to the surface, pores and sunspots are seen as dark structures in photospheric images. The major morphological feature distinguishing a sunspot from a pore is its penumbra, which is seen as radially aligned filaments surrounding the umbra. Pores have a similar structure as the umbrae of sunspots, but are generally smaller in size and have weaker magnetic fields \citep{1996A&A...316..229K}, although values of more than \SI{2000}{\gauss} can be observed \citep[e.g.][]{2016MNRAS.460.1476Q}. With increasing distance from the center of a pore, the magnetic fields get weaker and more inclined with inclination angles reaching \SI{40}{\degree} \citep{1996A&A...316..229K} or even higher  \citep{1998A&A...333..305S}. Magnetic fields in the umbra of sunspots are stronger with field strength values of more than \SI{3000}{\gauss}, but also generally get more inclined with increasing distance from the center \citep{2011LRSP....8....4B}. The largest inclination values in sunspots, however, are found in the penumbra with approximately \SI{70}{\degree} at the outer edge \citep{2011LRSP....8....4B}. Data from high resolution spectropolarimetric observations  \citep[e.g.][]{2008ApJ...689L..69S,2013A&A...553A..63S,2013A&A...557A..25T,2016A&A...596A...2B} show that the photospheric magnetic field in the penumbra follows a filamentary pattern: The magnetic field vector alternates between being stronger and less inclined in the spines and being weaker and more inclined in the intraspines. From sophisticated inversions and interpretation of the net circular polarization, it is also found that a horizontal flow component associated with weaker magnetic field is embedded in a less inclined magnetic field component at rest \citep{2003A&A...403L..47B,2002A&A...381L..77S,2002A&A...393..305M}, as proposed in the models of the uncombed penumbra \citep{1993A&A...275..283S,1998A&A...337..897S}. The predominant flow signature in the penumbra is the horizontal Evershed flow \citep{1909MNRAS..69..454E}, which is directed radially outward. Individual filaments carrying this flow have also been observed to exhibit upflows on their heads and downflows at their tails, accompanied by upward pointing magnetic fields and downward pointing magnetic fields, respectively  \citep{2013A&A...557A..25T}. In the dark parts of the umbra (outside light bridges, umbral dots, etc.), vertical velocities are reduced to values below \SI{100}{\meter \per \second} \citep{2018A&A...617A..19L}. 

The magnetic structure of sunspots and pores change toward higher layers in the chromosphere. Recent spectropolarimetric observations of chromospheric lines \citep[e.g.][]{2017A&A...604A..98J} provide evidence of a chromospheric canopy in the form of horizontal magnetic fields above the penumbra of sunspots, although the imprints of the spine/intraspine structure are seen up to at least the middle chromosphere \citep{2019ApJ...873..126M}. The region above and beyond the photospheric penumbra is also referred to as the `superpenumbra', showing radial inflows, that is, flow directions opposite of the photospheric Evershed flow \cite[e.g.][]{2017A&A...604A..98J}. 

In order to understand the evolution of pores and sunspots, processes at different heights and their interaction need to be taken into account. Sunspots evolve from pores (then called protospots) that accumulate further magnetic flux and at some point a penumbra is forming. \citet{1996A&A...316..229K} estimated a value of $\approx$\SI{3e20}{\maxwell} \citep[in rough agreement with the model by][]{1995MNRAS.273..491R} that is required for a pore to start penumbra formation, while newer studies \citep{2014SoPh..289.1143T,2015ApJ...811...49C} suggest a more fluent transition at slightly higher flux values. Several studies additionally suggested that a chromospheric canopy plays a major role in the formation of a penumbra to form around a protospot. \citet{2012ApJ...747L..18S} observed an annular zone in chromospheric Ca II H images before filaments in the photosphere were visible. Further observational evidence came from \citet{2021A&A...653A..93M}, who found that a penumbra formed in regions where magnetic field extrapolations indicated the presence of a chromospheric canopy.  In addition, Lindner et al. (2023, [proper reference pending]) showed spectropolarimetric data of a penumbra that formed around a sunspot, except in a region in which the chromospheric canopy was disturbed. The important role of a chromospheric canopy is also supported by numerical simulations, in which a sunspot develops a penumbra or not depending on the degree at which magnetic fields are forced to be inclined at the top of the simulation box \citep{2012ApJ...750...62R}. 

Solely in the photosphere, the formation of a penumbra has been observed, e.g. by \citet{2010A&A...512L...1S,2016ApJ...825...75M,2013ApJ...769L..18L}, revealing that the penumbra forms in segments and usually starts forming on the side facing the opposite polarity. Just before the formation of the penumbra, inward (counter-Evershed) flows have been detected by \citet{2012ASPC..455...61S}, which inverse their direction after the penumbral filaments have formed. \cite{2018ApJ...857...21L} reported on a special case of penumbra formation that is accompanied by photospheric blueshifts only. 

During the evolution of sunspots, signs of magnetic reconnection were also found outside flaring regions.  \citet{2007Sci...318.1594K} discovered elongated brightenings in chromospheric Ca II H images, which were called penumbral microjets (PMJs). They also leave spectral signatures in other chromospheric lines \citep{2020A&A...638A..63D}. Recent studies of PMJs with polarimetric data \citep{2019ApJ...870...88E,2020A&A...638A..63D} suggested that PJMs originate from magnetic reconnection occurring above sheared magnetic field configurations (e.g. boundaries between spines and intraspines) in the photosphere. This leads to an increase in low chromospheric/high photospheric temperatures. The lack of strong velocities in the vicinity of PMJs supports the scenario of heat transport by conduction with a perturbation front propagating along pre-existing fibrils. \cite{2020A&A...642A.128S} analyzed high-cadence spectropolarimetric data and found that in most cases, inclination and field strength values are increased during the phase of a PMJ. However, it still remains difficult to locate a certain event in the photosphere as the source of a PMJ \citep{2019A&A...626A..62R}. Signs of a reconnection event close to a pore in a flux emergence region were found by \citet{2021A&A...647A.188D}. The authors suggested that new emerging magnetic fields interacted with pre-existing fields, leading to magnetic reconnection at the location where flux cancellation happened in the photosphere and a chromospheric brightening was visible. The model atmosphere showed bipolar flow fields and a chromospheric temperature increase.

The transition from a pore to a sunspot does not always have to happen completely and at once. Intermediate stages exist and some authors \citep{2012ApJ...746L..13S} also question whether a naked sunspot/pore that shows no filaments in intensity, but has a magnetic surrounding extending significantly beyond the visible boundaries, should be classified as a sunspot or not. In rare cases, single transient filaments around pores also form without a whole segment of penumbra. \citet{2008ApJ...676..698B} analyzed `finger-like' inhomogeneities at the border of an umbra, whose penumbra had decayed before. These structures were interpreted to be linked to the previously existing penumbra and showed inclined magnetic fields and upflows, but left no imprint in intensity maps.  \citet{2016MNRAS.462L..93B} observed small filamentary structures in which inclined fields were measured, together with Y-shaped dark cores. \citet{2014ApJ...796...77W} observed a `transitory' penumbral segment that formed at the border of a pore, but decayed after 5 hours. However, no spectropolarimetric data was available. In this work, we are analyzing the decay of a transient filament at the border of a pore with high-resolution spectropolarimetric inversions from photospheric and chromospheric lines. We are also aiming toward understanding the possible connection of a chromospheric response with the photospheric evolution.

\section{Data and methods}
\label{data}
\subsection{Observations}
\begin{figure}[h!]
\centering
	\includegraphics[width=8.8cm]{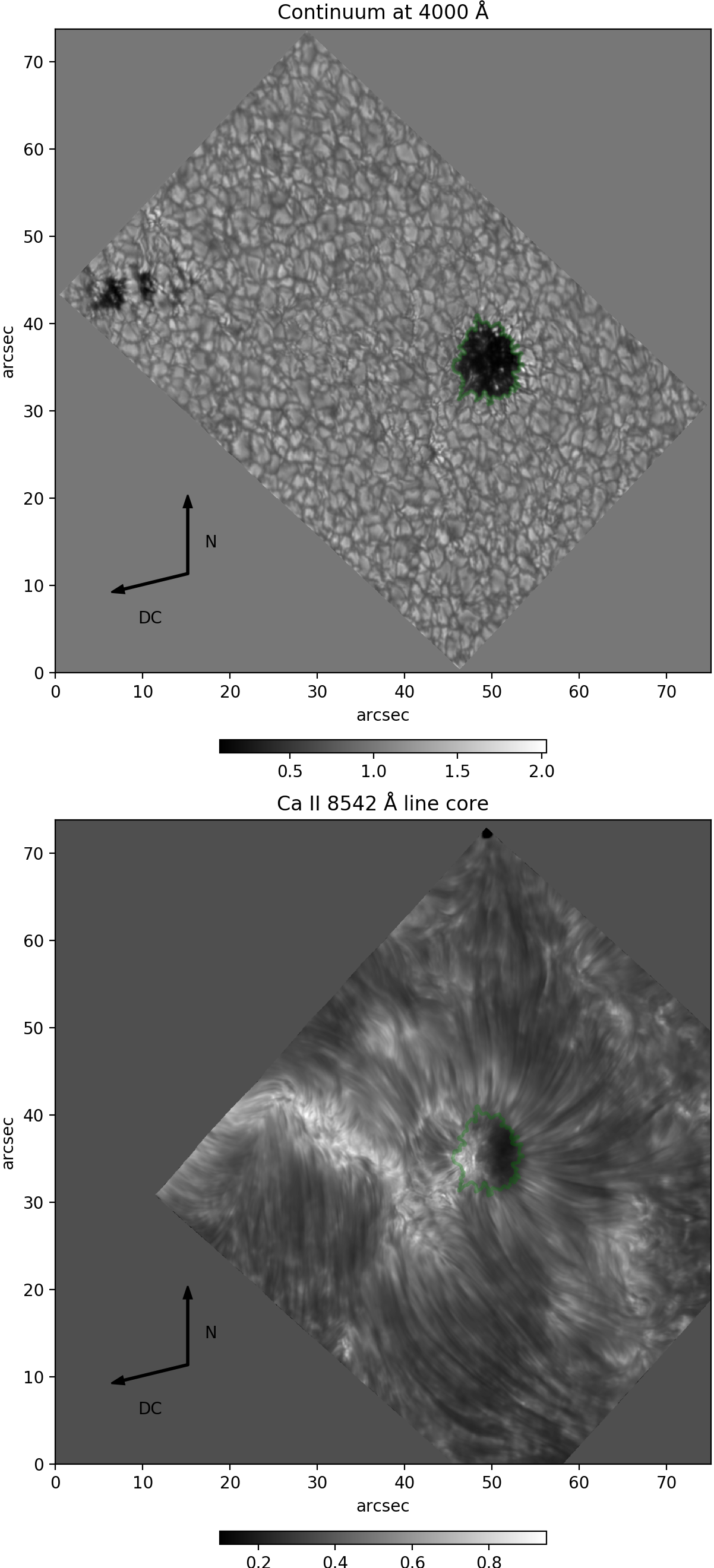}
     \caption{Full FOV maps showing an image at the continuum wavelength point for CHROMIS at \SI{4000}{\angstrom} (top) and one image at wavelength in the core of the \ion{Ca}{II}~8542~\SI{}{\angstrom} line for CRISP (bottom). The intensities were normalized, separately for each line, to the (quasi) continuum value averaged over a quiet sun region at the bottom of the FOV. The time step shown here is the first one at 10:33UT. The green contour outlining the pore was produced from the CHROMIS image at the continuum wavelength point at \SI{4000}{\angstrom}.} 
     \label{overview}
\end{figure}

The data used in this study were taken at the Swedish Solar Telescope (SST) \citep{2003SPIE.4853..341S} during a campaign funded by the SOLARNET Transnational Access and Service Programme. It includes the Active Region (AR) NOAA 12739 with a large pore in the leading polarity and several small ones in the following polarity (see Fig.~\ref{overview}, top panel). Data was taken on 2019-04-19, where seeing conditions were variable but allowed for a short sequence of data recording between 10:33~UT and 10:46~UT. At that time, NOAA 12739 was located  at the helioprojected cartesian latitude/longitude of approximately [157~arcsec,~615~arcsec], corresponding to a heliocentric angle $\theta$ of \SI{41}{\degree} or $\mu = \cos (\theta)$~=~0.75. The CRISP instrument \citep{2006A&A...447.1111S,2008ApJ...689L..69S} recorded the \ion{Fe}{I}~6173~\SI{}{\angstrom} and the \ion{Ca}{II}~8542~\SI{}{\angstrom} line with full polarimetry \citep{2021AJ....161...89D} and a full cadence of \SI{22}{\second}. The \ion{Fe}{I}~6173~\SI{}{\angstrom} line was spectrally sampled with a wavelength spacing of \SI{35}{\milli \angstrom} close to the core and \SI{50}{\milli \angstrom} in its wings. For the \ion{Ca}{II}~8542~\SI{}{\angstrom} line, \SI{75}{\milli \angstrom} and \SI{125}{\milli \angstrom} were used. The CHROMIS instrument \citep[]{2017psio.confE..85S} recorded the \ion{Ca}{II}~K~3934~\SI{}{\angstrom} line in spectroscopy mode with a cadence of \SI{8}{\second}. An equidistant wavelength sampling with a spacing of \SI{65}{\milli \angstrom} was used for the entire line scan. In addition, CHROMIS recorded data in the continuum at a wavelength of \SI{4000}{\angstrom}. Photospheric and chromospheric images of the full field of view (FOV) of the respective instruments are shown in Fig.~\ref{overview}.

\subsection{Processing}
The data from both instrument was processed with the SSTRED pipeline \citep{2015A&A...573A..40D,2021A&A...653A..68L}, including a MOMFBD reconstruction \citep{2005SoPh..228..191V}. Additionally, a spatial alignment was performed, where the CHROMIS data was taken as a  reference. The image scale of the CRISP data originally is 0.0592~arcsec/pixel, but was resampled to the CHROMIS image scale of 0.0379~arcsec/pixel. For the temporal alignment, however, we chose CRISP as a reference: For each of the 35 CRISP frames, we chose the CHROMIS frame that is closest in time to the \ion{Ca}{II}~8542~\SI{}{\angstrom} line scan. The intensity scaling and the wavelength shift of each line were calibrated using quiet sun data from disk center (recorded approximately \SI{30}{\minute} after the observations), which were compared to the solar atlas \citep{1984SoPh...90..205N}. For the wavelength calibration, we additionally corrected for the solar rotation introducing a wavelength shift between the disk center data and the data from the Active Region. The rotational shift was determined using velocity maps from the Helioseismic and Magnetic Imager (HMI) \citep{2012SoPh..275..229S} onboard the Solar Dynamics Observatory (SDO) \citep{2012SoPh..275....3P}, so that differential rotation effects are also included. The \ion{Fe}{I}~6173~\SI{}{\angstrom} line is suitable for determining the rotational shifts, because for  $\mu$~>~0.5, convective blueshifts do not change with the heliocentric angle by more than \SI{50}{\meter \per \second} \citep{2019A&A...624A..57L}.

\subsection{Evolution of NOAA AR 12739}
\label{hmi}
Before our observations with the SST started, the pore had been visible on the sun for at least two days. HMI continuum images (not shown in this paper) show that the pore accumulated flux, but no clear penumbra was visible. Two days after the SST observations, the pore was not recognizable any more because it approached the solar limb. During the five days when it was visible, the pore stayed remarkably stable while, in the following polarity, multiple small pores formed and decayed. The SHARP data series \citep{2014SoPh..289.3549B} produces a magnetic flux value (header keyword `USFLUX') and a de-projected area covered by active pixels (header keyword `AREA\_ACR'). For this Active Region on 2019-04-19 at 10:36UT, a flux value of \SI{1.6e21}{\maxwell} and an area of \SI{55.8}{\mh} (millionth of the solar hemisphere) was calculated. In the distribution of pores and sunspots from \citet[Fig. 2]{2014SoPh..289.1143T} and \citet[Fig. 2]{2015ApJ...811...49C}, a pore/sunspot with these two values would be categorized as a transitional sunspot. A direct comparison to their distribution, however, is not accurate because they used special methods to calculate flux and area values.

\subsection{VFISV Inversions}
\label{vfisv_methods}
\begin{figure*}[h!]
\centering
	\includegraphics[width=18cm]{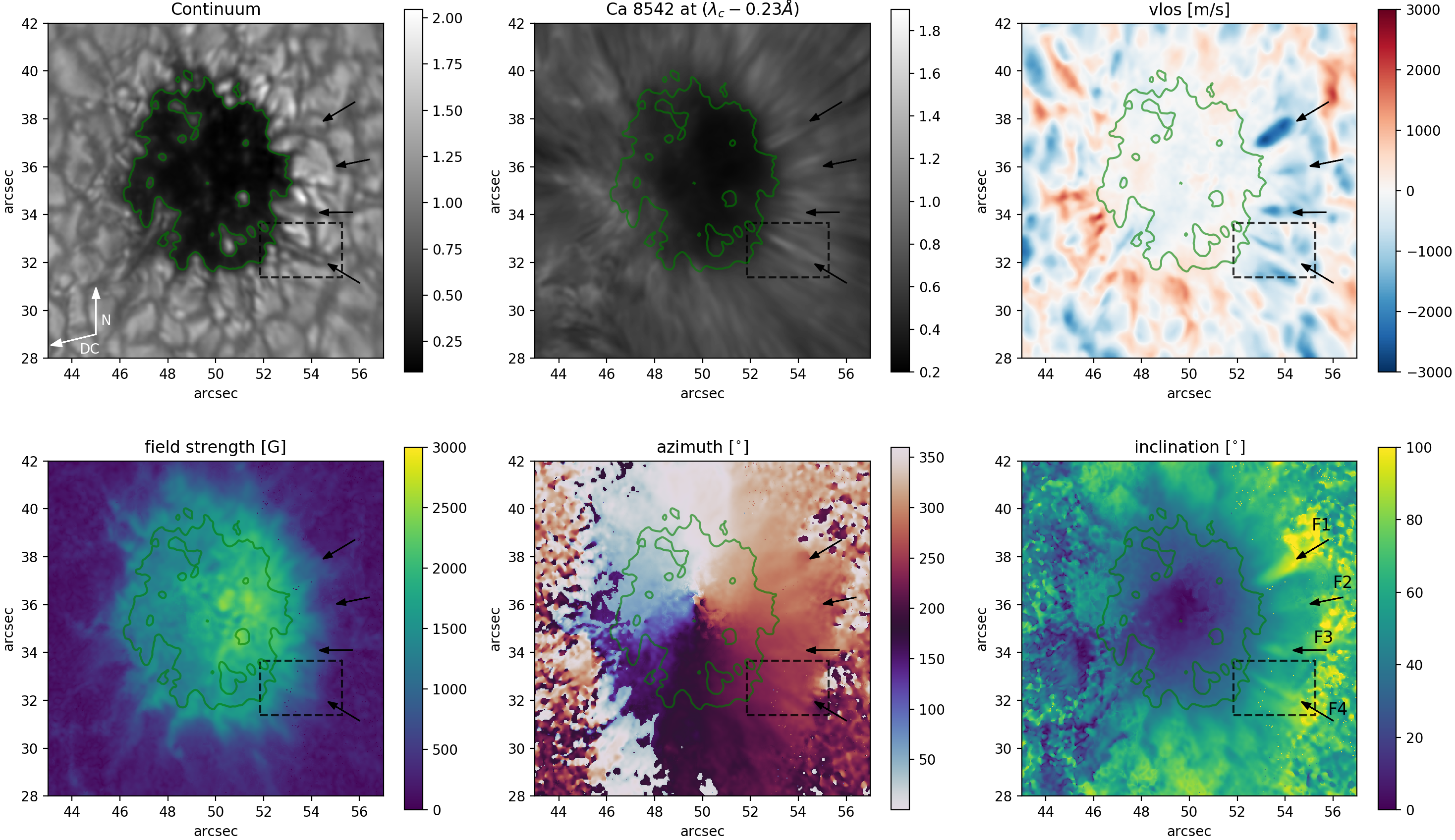}
     \caption{Maps showing continuum (from the CHROMIS instrument at \SI{4000}{\angstrom}) on the top left, the blue wing intensity from the \ion{Ca}{II}~8542~\SI{}{\angstrom} line (from the CRISP instrument) at the top center, and the VFISV inversion results LOS velocity (top right), magnetic field strength (bottom left), LRF azimuth angle (bottom center) and LRF inclination angle of the magnetic field (bottom right). The value range of the inclination map has been clipped for better visibility. Time step: $t_{\mathrm{obs}}$ = 0 sec. The green contour outlining the pore was produced from the CHROMIS image at the continuum wavelength point at \SI{4000}{\angstrom} and is overplotted on the other panels.}
     \label{vfisvmaps}
\end{figure*}

VFISV \citep{2011SoPh..273..267B} is a Milne-Eddington inversion code optimized for computational efficiency and speed. We used it to obtain velocity maps and magnetic field vector maps from the full-Stokes spectra of the \ion{Fe}{I}~6173~\SI{}{\angstrom} line for the full CRISP FOV and all time steps. The code is available on the the GitLab platform\footnote{https://gitlab.leibniz-kis.de/borrero/vfisv\_fpi}, from where we downloaded it on 2022-04-07. We used a configuration in which the following parameters were inverted: Center to continuum
absorption coefficient, magnetic field strength, inclination and azimuth, damping parameter, Doppler width, line of sight (LOS) velocity, continuum source function value, and the gradient of the source function. The filling factor was kept at a constant value of 1. The magnetic field vector maps were transferred from the LOS reference frame to the local reference frame (LRF). The \SI{180}{\degree} ambiguity in the maps of the LOS azimuth angle of the magnetic field was resolved in a pragmatic way: As the pore was located at a heliocentric angle of \SI{41}{\degree} during our observations, the two solutions from the LOS azimuthal ambiguity lead to two solutions in the LRF that have different inclination values. We attempted to resolve the ambiguity by selecting the more vertical solution for each pixel. As shown in Fig.~\ref{vfisvmaps}, this resulted in a smooth distribution of the LRF azimuth in the pore and its magnetic surroundings. Only in the bottom-left (south-eastern) side of the sub-FOV, a region with discontinuities is seen. However, this region has a complex magnetic structure in which applying a manual disambiguation would also be problematic. In the following, we are not analyzing this region, so we proceeded with this disambiguation method. The velocity maps were calibrated by adding a constant offset to all time steps such that the average velocity inside the pore and over all time steps is zero. Maps showing atmospheric parameters for the first time step, cropped around the pore, are shown in Fig.~\ref{vfisvmaps}. 

\section{Analysis and results}
In this section, filamentary structures identified at the outskirts of the pore in inclination maps from the VFISV inversion  are analyzed. For one of them, an elongated brightening as seen in images in the blue wing of the \ion{Ca}{II}~8542~\SI{}{\angstrom} line is observed during its decay. The evolution of this event is studied with both the results from the VFISV inversion of the \ion{Fe}{I}~6173~\SI{}{\angstrom} line and  with results from the  STiC non-LTE inversion code \citep{2019A&A...623A..74D} of one time step taking the \ion{Fe}{I}~6173~\SI{}{\angstrom}, the \ion{Ca}{II}~8542~\SI{}{\angstrom} and the \ion{Ca}{II}~K~3934~\SI{}{\angstrom} lines into account.

\subsection{Characterization of transient filaments}
Maps displaying the results from the VFISV inversions of the photospheric \ion{Fe}{I}~6173~\SI{}{\angstrom} line and the first time step of the observation are shown in Fig.~\ref{vfisvmaps}. In the map showing the LRF inclination angle of the magnetic field, multiple elongated structures with increased inclination values are seen on the right (western) side of the pore. We call them transient filaments, since they are radially oriented, adjacent to the pore, but they are observed only for a limited time span of our observations. HMI data additionally show that these filaments do not develop into a penumbra. They are marked by arrows in Fig.~\ref{vfisvmaps} and are named from F1 to F4, from top to bottom. The filaments differ in size and in inclination values with F1 being the most prominent one and F2 being very faint to recognize. 

The LOS velocity maps reveal that the transient filaments are accompanied by blueshifts (flows approaching the observer). In continuum intensity images, only filament F4 clearly shows a structure that can be linked to the respective transient filament. In this case, an elongated bright structure with a shape similar to the filament seen in the inclination map is clearly visible at the same location. For the other filaments, the situation is different. At the location of F1, a short and rather thick bright structure is visible, that could also be a small granule. At the location  of F2, only a large granule is visible. In location F3, a dark structure is visible, that could have a similar shape as the filament seen in the inclination map, but could also be a regular intergranular lane. We also note that in general, blueshifted structures are present more on the limb side of the pore and red-shifted structures more on the disk-center side. 

In velocity maps from HMI at the time of the ground-based observation (not shown here) we were able to identify the large blueshift patch belonging to filament F4. In addition, several smaller blueshifted structures are seen at the location of the other filaments, but the limited spatial resolution of HMI does not allow for an explicit assigning. Scanning through all HMI velocity maps from that day, we note that such blueshifted patches appear and disappear on the West side of the pore for the whole day.

In addition, a chromospheric image in the blue wing of the \ion{Ca}{II}~8542~\SI{}{\angstrom} line at a wavelength of $\lambda = \lambda_c - \SI{0.23}{\angstrom}$, with $\lambda_c$ being the wavelength of the line core, shows a clear elongated brightening at the location of filament F4 (see Fig.~\ref{vfisvmaps}, top center). 

In the following, we are focusing the analysis on filament F4, because it clearly shows a filamentary feature in inclination maps, velocity maps and continuum intensity maps. Additionally, it shows an elongated brightening in the blue wing of the \ion{Ca}{II}~8542~\SI{}{\angstrom} line, which is analyzed further in Sect.~\ref{cabright}.

\subsection{Structure and temporal evolution of filament F4}
\begin{figure}[h!]
\centering
	\includegraphics[width=8.8cm]{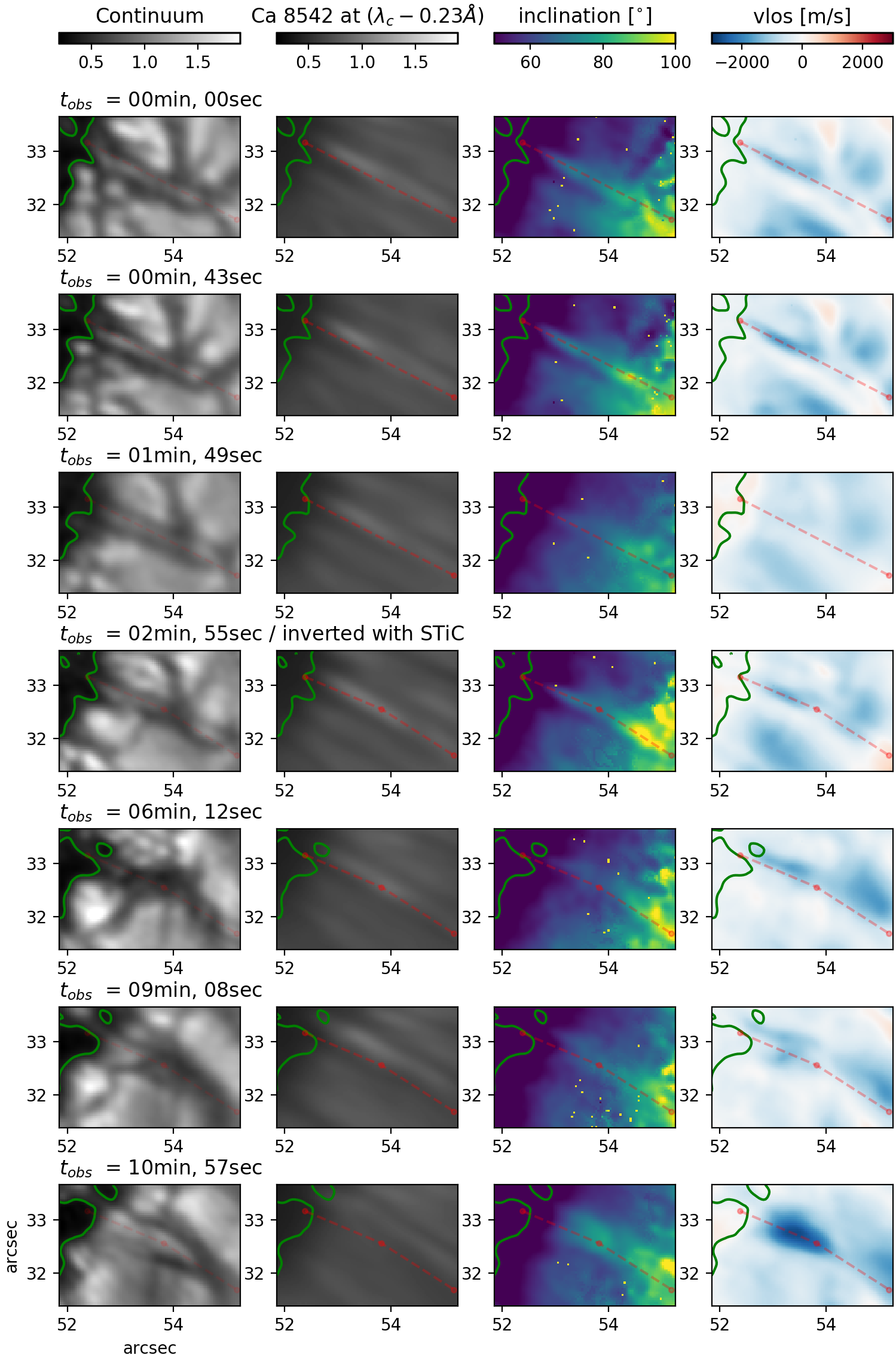}
     \caption{Cropped maps that show the evolution of filament F4. LRF inclination and velocity maps show the result from the VFISV inversions of the \ion{Fe}{I}~6173~\SI{}{\angstrom} line. Time counting starts at the starting point of our observations at 10:33~UT. The value range of the inclination map was set to [40,100] for better visibility in this small region. The green contour outlining the pore was produced from the CHROMIS image at the continuum wavelength point at \SI{4000}{\angstrom} and is overplotted on the other panels.}
     \label{vfisv_fil}
\end{figure}

Detail maps of filament F4 for selected time steps are shown in Fig.~\ref{vfisv_fil}. The continuum maps were taken with the CHROMIS instrument, and the LRF inclination and velocity maps are from the VFISV inversion results of the \ion{Fe}{I}~6173~\SI{}{\angstrom} line. Additionally, images in the blue wing of the \ion{Ca}{II}~8542~\SI{}{\angstrom} at $\lambda_c - \SI{0.23}{\angstrom}$ are displayed. Red dashed lines are over-plotted to show the axis of the filament. These lines were manually determined from the maps of the magnetic field inclination.

In the first three time steps displayed in Fig.~\ref{vfisv_fil}, F4 remains stable in continuum maps and in inclination maps. In continuum maps, it is visible as a filamentary bright feature surrounded by dark lanes with a length of approximately \SI{2}{\arcsec}. In the images of the \ion{Ca}{II}~8542~\SI{}{\angstrom} blue wing, an elongated brightening is visible. It is oriented parallel to F4, located at the left (eastern) half of the filament and is slightly north of the filament's central axis. In the inclination maps of the first three time steps, the inclination values along the filament are approximately constant at \SI{70}{\degree}. Only the part close to the pore shows smaller (more vertical) values. This changes at $t_{\mathrm{obs}}$=02min 55sec. In the continuum, a dark structure develops in the center of the filament, splitting the filament into two: A left part and a right part. In the inclination maps, the right part shows further increased values above \SI{90}{\degree}, that is, magnetic fields of opposite polarity. In addition, the axis of the filament is not straight anymore, which we tried to follow by plotting a red dashed line with a kink in the center. The brightening in the \ion{Ca}{II}~8542~\SI{}{\angstrom} blue wing has extended slightly, but no kink is noticeable. At $t_{\mathrm{obs}}$=6min 12sec, the filament is not visible as a bright elongated filament in the continuum map anymore. Instead, a dark extended structure is seen. The brightening in the \ion{Ca}{II}~8542~\SI{}{\angstrom} blue wing shows decreased values. In the inclination map, the filament is not clearly recognizable anymore. At $t_{\mathrm{obs}}$=09min 08sec and $t_{\mathrm{obs}}$=10min57sec, the extended dark structure is diminishing and new dark lanes form. The \ion{Ca}{II}~8542~\SI{}{\angstrom} blue wing brightening is not visible any more. In the inclination map, a new elliptically shaped structure with increased values is visible at $t_{\mathrm{obs}}$=10min 57sec, on the right part of the former filament. 

Overall, the velocity maps do not follow the break-up of the filament so clearly. In the first time steps, blueshifts are visible in the left part of the filament, while the right part shows a more diffuse structure. At this position on the solar disk, blueshifts represent an upflow, or also a horizontal motion toward disk center (left of the FOV.) After the break-up of the filament axis, blueshifts are present in the left and right part of the filament. The highest values, however, are seen at $t_{\mathrm{obs}}$=10min 57sec at the location of the newly forming elliptically shaped structure seen in both velocity and inclination maps.  

\begin{figure}[h!]
\centering
	\includegraphics[width=8.8cm]{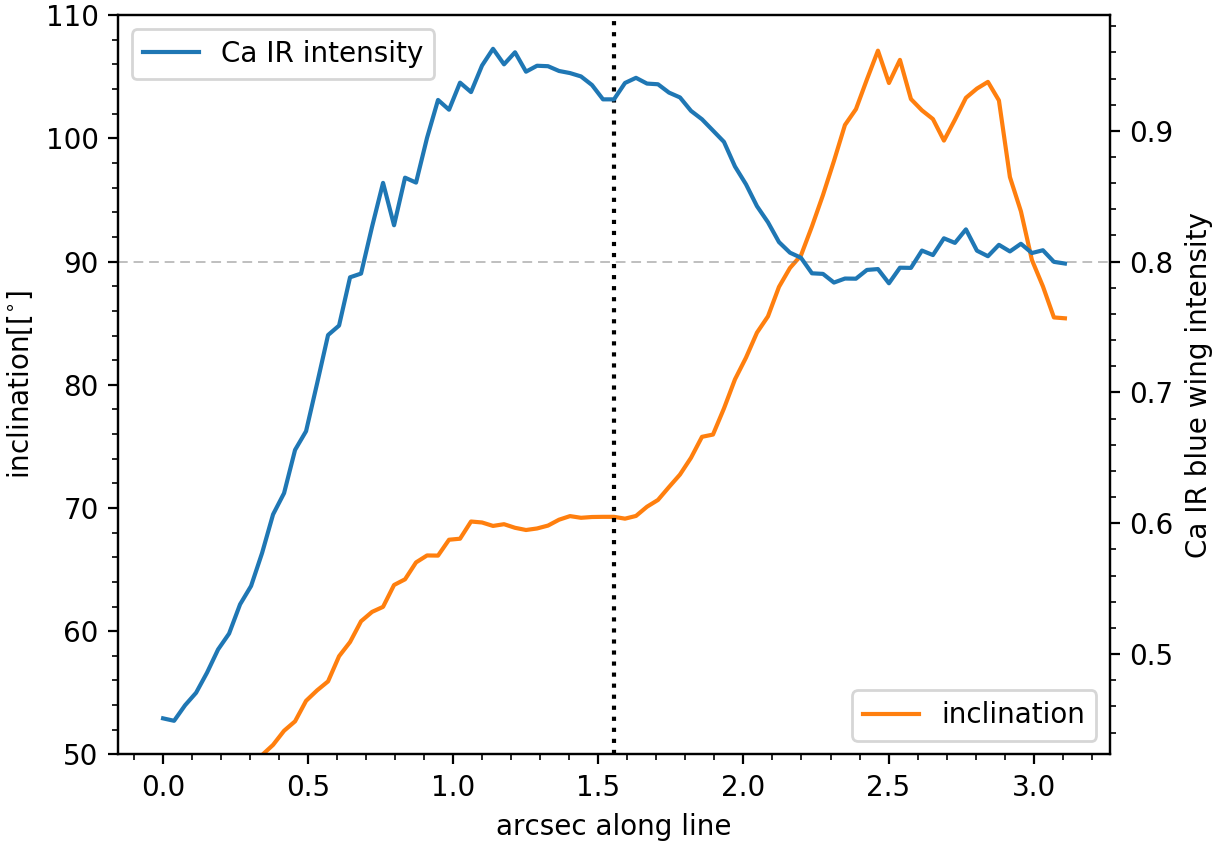}
     \caption{LRF inclination (orange line) and \ion{Ca}{II}~8542~\SI{}{\angstrom} blue wing intensity (blue line, second axis on the right) along the red dashed line plotted in Fig.~\ref{vfisv_fil}. The value range of the inclination axis is cropped at \SI{50}{\degree} for better visibility of the distribution. The \ion{Ca}{II}~8542~\SI{}{\angstrom} intensities were normalized to a an average over the full FOV of a quasi continuum wavelength point. Time:~$t_{\mathrm{obs}}$=02min 55sec} 
     \label{incalong}
\end{figure}

In order to show a quantified analysis of the properties of the filament F4, the inclination values along the red dashed line (shown in Fig.~\ref{vfisv_fil}) at $t_{\mathrm{obs}}$=02min 55sec are plotted as an orange line in Fig.~\ref{incalong}. At the location of the kink of the red dashed line (see Fig.~\ref{vfisv_fil}), a vertical black dashed line is plotted. On the left part of the filament (at the pore boundary), inclination values below \SI{50}{\degree} are present. The values increase until reaching a plateau of \SI{70}{\degree} at the center of the filament. At the right part of the filament, the inclination values are exceeding \SI{90}{\degree}. This means that in the outer foot point of filament F4, the magnetic field reversed polarity. On a second axis in Fig.~\ref{incalong}, the intensity value of the blue wing of the \ion{Ca}{II}~8542~\SI{}{\angstrom} line at $\lambda_c - \SI{0.23}{\angstrom}$ is plotted along the red dashed line. The values reach almost the quasi continuum values coinciding with the \SI{70}{\degree} inclination plateau and peak slightly to the left of the kink of the red dashed line. 

\subsection{Temporal characterization of the \ion{Ca}{II}~8542~\SI{}{\angstrom} brightening}
\label{cabright}
\begin{figure}[h!]
\centering
	\includegraphics[width=8.8cm]{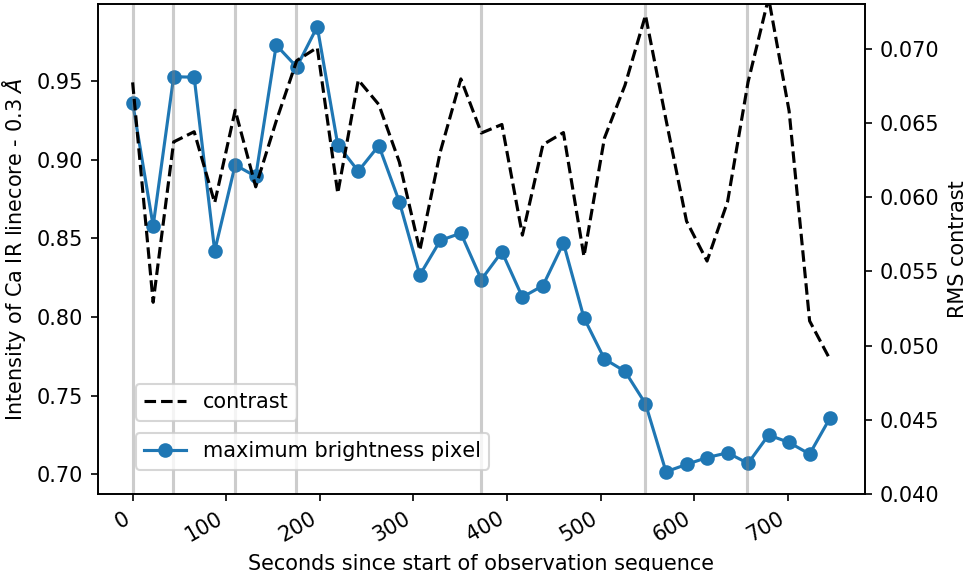}
     \caption{Light curve of the maximum brightness pixel and seeing in terms of RMS contrast of a quiet subfield. Gray vertical lines correspond to the time steps at which maps are shown in Fig.~\ref{vfisv_fil}} 
     \label{lightcurve}
\end{figure}

The brightening in the images of the blue wing of the \ion{Ca}{II}~8542~\SI{}{\angstrom} at $\lambda_c - \SI{0.23}{\angstrom}$ is present already in the first time steps of the observations (cf.~Fig.~\ref{vfisv_fil}). In order to characterize the temporal evolution of the brightening during the ground-based observations, Fig.~\ref{lightcurve} shows a lightcurve of the pixel that had the maximum brightness value over the full time series within the cropped region shown in Fig.~\ref{vfisv_fil}. In general, blurring effects from changes in seeing conditions can influence maximum values in the intensity images. Therefore, the standard deviation over a quiet region of the image was calculated at the same wavelength point (to ensure simultaneity). This value serves as a measure of the root mean square (RMS) contrast and therefore the seeing conditions and is plotted on a second axis. Until approximately \SI{240}{\second}, the intensity values alternate around approximately 0.9 and seem to follow the evolution of the RMS contrast. After \SI{240}{\second}, the intensity values decrease below 0.85 (below 0.75 after \SI{300}{\second}), while the RMS contrast continues to alternate around approximately 0.06. The \ion{Ca}{II}~8542~\SI{}{\angstrom} blue wing brightness therefore starts to diminish shortly (within one minute) after the break-up of the filament F4 seen in the photospheric inclination maps and continuum images.

\subsection{Spectral characterization of the \ion{Ca}{II}~8542~\SI{}{\angstrom} brightening}\label{spectral_characterization}
\begin{figure*}[h!]
\centering
	\includegraphics[width=18cm]{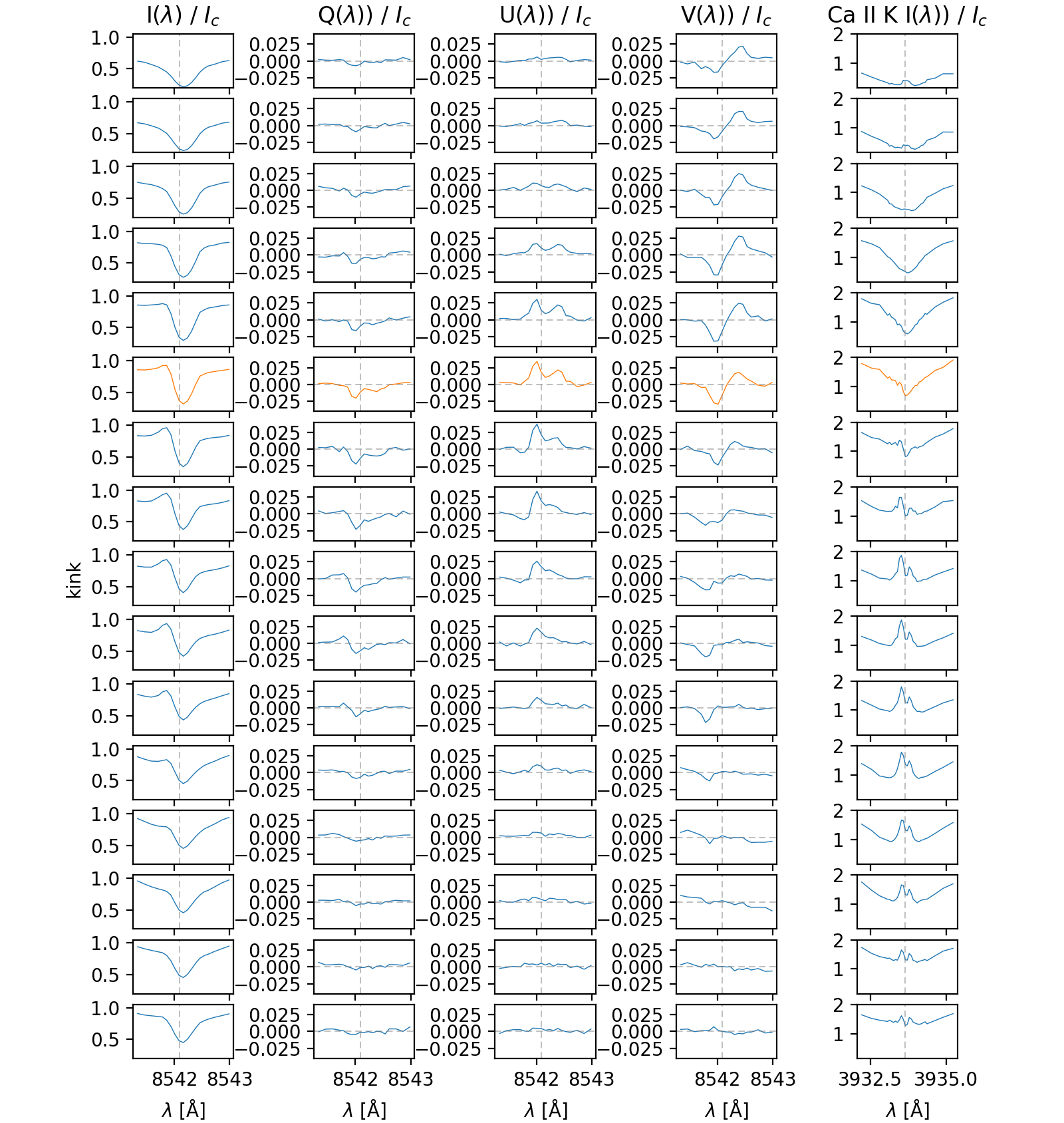}
     \caption{Column one to column four: Stokes-$I$, Stokes-$Q$, Stokes-$U$ and Stokes-$V$ of the \ion{Ca}{II}~8542~\SI{}{\angstrom} line. Column five (on the right): Stokes-$I$ of the \ion{Ca}{II}~K~3934~\SI{}{\angstrom}. The rows from top to bottom show spectra of pixels along the red dashed line shown in Fig.~\ref{vfisv_fil}. Every fifth pixel is shown here. The dashed vertical line depicts the laboratory wavelength of the line center. The spectra of the pixel at the kink of the red dashed line are shown in row 9 (labeled with the word `kink' on the left side of the plot). All spectra are from the time step at $t_{\mathrm{obs}}$=02min 55sec.} 
     \label{spec_along}
\end{figure*}

A brightening in the wing of a spectral line can have different spectral origins. Therefore, the spectra of the time step at $t_{\mathrm{obs}}$=02min 55sec are analyzed along the red dashed line plotted in Fig.~\ref{vfisv_fil}. Figure~\ref{spec_along} shows spectra from every fifth pixel along the red dashed line, starting from the left (closest to the pore). For the \ion{Ca}{II}~8542~\SI{}{\angstrom} line, all Stokes profiles are plotted and additionally the Stokes-$I$ profile of the \ion{Ca}{II}~K~3934~\SI{}{\angstrom} line observed with the CHROMIS instrument are plotted. The Stokes-$I$ profiles of the \ion{Ca}{II}~8542~\SI{}{\angstrom} line on the left part of the red dashed line (top rows of Fig.~\ref{spec_along}) are not completely symmetric, but do not show emission peaks. The \ion{Ca}{II}~K~3934~\SI{}{\angstrom} line shows a small emission peak in the line center for some of the first rows. Starting from row five, and more clearly in row six (in orange color), a peak is appearing in the blue wing of the \ion{Ca}{II}~8542~\SI{}{\angstrom}. The brightening in the images is therefore not caused by an overall increase in intensity over the full line, but is caused by a blue wing peak (hereinafter referred to as BWP). The BWP is standing out most clearly close to the kink of the red dashed line (labeled with the word `kink' in Fig.~\ref{spec_along}). In this region, the \ion{Ca}{II}~K~3934~\SI{}{\angstrom} profiles also show emission peaks around the line center that exceed the outer wings. Interestingly, however, there are pixels in which the BWP is visible in the  \ion{Ca}{II}~8542~\SI{}{\angstrom} line, but the \ion{Ca}{II}~K~3934~\SI{}{\angstrom} has a `quiet' shape (with none to almost no emission). The profiles in row six (orange color) represent such a pixel. On the right side of the red dashed line (bottom rows in Fig.~\ref{spec_along}), the BWP is not present anymore. The signals in Stokes-$Q$, Stokes-$U$ and Stokes-$V$ seem to be affected from the BWP: Especially Stokes-$U$ shows an increased amplitude of the blue peak for the rows where the BWP is present. For Stokes-$Q$ and Stokes-$V$, the amplitude increase is also present, but weaker than for Stokes-$U$. As the polarimetric Stokes parameters are produced by the LOS magnetic field vector, it is difficult to analyze the magnetic field from the spectra directly. We refer to the STiC inversion results presented in Sect.~\ref{stic} and the corresponding LRF magnetic field results.

\begin{figure}[h!]
\centering
	\includegraphics[width=7cm]{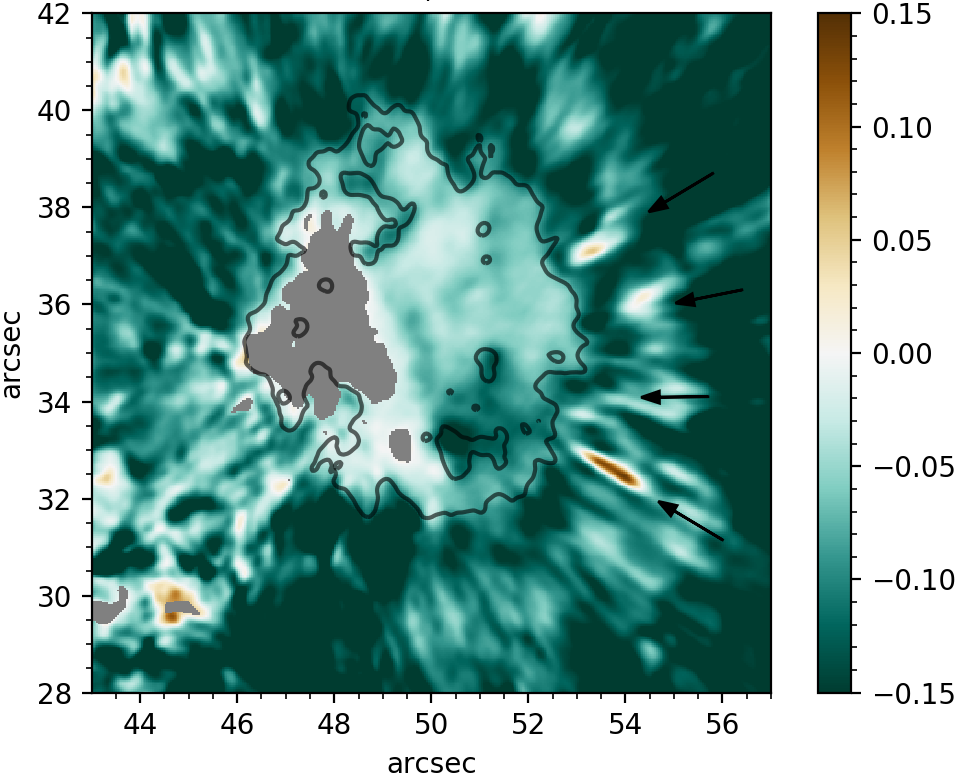}
     \caption{Maps showing the blue wing slope, as calculated from the intensity difference $I(\lambda_c - \SI{0.23}{\angstrom}) - I(\lambda_c - \SI{0.50}{\angstrom})$ for the \ion{Ca}{II}~8542~\SI{}{\angstrom} line. Pixels showing a strong line core emission are masked out in gray. Time step: $t_{\mathrm{obs}}$=02min 55sec. The black contour outlining the pore was produced from the CHROMIS image at the continuum wavelength point at \SI{4000}{\angstrom}.}
     \label{bspos}
\end{figure}

The presence of the BWP in the \ion{Ca}{II}~8542~\SI{}{\angstrom} line motivated to find a more sophisticated way to identify brightenings of this nature. Difference maps between the intensity at the wavelength corresponding to the BWP and the intensity at the quasi continuum were produced by calculating $I(\lambda_c - \SI{0.23}{\angstrom}) - I(\lambda_c - \SI{0.50}{\angstrom})$. For one time step, such a difference maps is shown in Fig.~\ref{bspos}. As shown by black vertical dashed lines in the example spectra displayed in Fig.~\ref{example}, this difference is positive for pixels with a positive blue wing slope (Stokes-$I$ increasing with wavelength) in this spectral region. As this can be the case not only for pixels showing a BWP, but also for pixels that show a strong line core emission, we masked out pixels with strong central emissions. They were identified by the condition of a line core value being higher than the value of the outer wing, that is, $I(\lambda_c) - I(\lambda_c - \SI{0.23}{\angstrom}) > 0$. These pixels are masked out with gray color in Fig.~\ref{bspos}. 

We manually verified that pixels showing a positive value in the blue wing slope maps and no central core emission, indeed, show a BWP. In particular, several elongated structures with positive blue wing slope values were identified not only close to filament F4, but also close to F1 and F2. These regions were smaller than the region at filament F4. A manual inspection confirmed that a BWP was present in these regions, but less pronounced.

\subsection{STiC inversion strategy}
\label{stic}
In order to further analyze the brightening in the  \ion{Ca}{II}~8542~\SI{}{\angstrom} blue wing and a possible connection to the co-spatial and co-temporal evolution of filament F4, we used the  STockholm Inversion Code \citep[STiC,][]{2019A&A...623A..74D,2016ApJ...830L..30D}. STiC\footnote{https://github.com/jaimedelacruz/stic} is a MPI-parallel non-LTE inversion code that utilizes a modified version of RH \citep{2001ApJ...557..389U} to solve the atom population densities assuming statistical equilibrium and plane-parallel geometry. It allows including partial redistribution effects of scattered photons \citep{2012A&A...543A.109L}, which is important e.g. for the synthesis of the \ion{Ca}{II}~K~3934~\SI{}{\angstrom} line. The radiative transport equation is solved using cubic Bezier solvers \citep{2013ApJ...764...33D}. The inversion engine of STiC includes an equation of state extracted from the SME code \citep{2017A&A...597A..16P}. The merit function (describing the differences between observational data and the current inversion fit) is adapted by adding a regularization term. The regularization can dampen oscillatory behaviors in the height stratification of atmospheric parameters in the case of using a high number of nodes \citep[see][]{2019A&A...623A..74D}. 

The main aim of the inversion with the STiC code was to fit the BWP of the \ion{Ca}{II}~8542~\SI{}{\angstrom} line, also for the complex pixels in which the \ion{Ca}{II}~K~3934~\SI{}{\angstrom} line is `quiet' (see, e.g. Fig.~\ref{spec_along}, orange row). The \ion{Fe}{i}~6173~\SI{}{\angstrom}, the  \ion{Ca}{II}~8542~\SI{}{\angstrom} and the \ion{Ca}{II}~K~3934~\SI{}{\angstrom} (only Stokes-$I$ available) lines were inverted simultaneously in one atmospheric model. As spectral information from a broad range of formation heights is taken into account, the resulting stratification of atmospheric parameters are also valid for a broad  height range. Four iteration cycles were used and the result of the previous cycle was used as an initial model for the following cycle. The initial model for the first cycle was based on the FALC atmospheric model \citep{1993ApJ...406..319F}, with the magnetic field vector calculated by using the weak-field-approximation. The number of nodes used for the different cycles is listed in Table~\ref{invstrat}. 

\begin{table}[h]
\caption{Number of nodes used in the multiple cycles of the STiC inversion.}
\label{invstrat}
\begin{tabular}{ |c|c c c c c c| }
\hline
cycle & temp & vlos & blong & bhor & azi & polarimetry \\ 
\hline
1 & 7 & 5 & (2) & (2) & (1) & low weight  \\
2 & 8\tablefootmark{*} & 5 & (2) & (2) & (1) & low weight  \\
\rowcolor{lightgray} 3 & 8\tablefootmark{*} & 5\tablefootmark{**} & (2) & (2) & (1) & low weight \\
\rowcolor{lightgray} 4 & 8\tablefootmark{*} & 5 & 2 & 2 & 1 & yes  \\
\hline
\end{tabular}
\tablefoot{
\tablefoottext{*}{manually chosen node locations and}
\tablefoottext{**}{five additional velocity stratifications were used as initial model.}
}
\end{table}

In the first two cycles, only the \ion{Fe}{i}~6173~\SI{}{\angstrom} and the \ion{Ca}{II}~8542~\SI{}{\angstrom} lines were inverted. \ion{Ca}{II}~K~3934~\SI{}{\angstrom} was added in cycle three (colored gray in Table.~\ref{invstrat}). Nodes were in general placed automatically by the code, which uses an equidistant node placement. Only for temperature, the nodes were not placed equidistantly from the second cycle onward. Instead, the location of the nodes was chosen manually (at $\log\tau=$[-6.5, -5.0, -4.25, -3.5, -2.75, -2.0, -1.25, 0.25]$)$ to set a more dense placement of nodes in the region around $\log\tau=-3.5$. All optical depth values $\tau$ used here and in the following refer to the optical depth at a wavelength of \SI{500}{\nano \meter}. Forward modeling tests (not shown here) had yielded that a temperature increase in this region can produce a BWP. The weights for Stokes Q, Stokes U and Stokes V were 15 times smaller than for Stokes I for the first three cycles, to give more weight on the Stokes I, which includes the BWP. In the last cycle, weights on polarimetry were given in the same order as for Stokes I. 

In cycle three, five different artificially defined velocity stratifications were used in addition to the result from cycle two. The additional velocity stratifications represented (1) a linearly increasing velocity, (2) a linearly decreasing velocity, (3) a constant negative velocity, (4) a constant velocity at the value of zero, and (5) a constant positive velocity \citep{2022A&A...664A...8M}. For each of the velocity inputs (six in total), a separate sub-cycle was ran to avoid for the code to find only a shallow local minimum in the multi-dimensional $\chi^2$ surface. $\chi^2$ describes the difference between the observed spectra and the resulting inversion fit. From these six sub-cycles, the result with the lowest $\chi^2$ value was chosen as the result of cycle three. This technique is useful especially when inverting the \ion{Ca}{II}~K~3934~\SI{}{\angstrom}, where cross-talk between the multiple peaks in the line core is possible. 

A full inversion run of the chosen region with 60 by 90 pixels including all cycles required approximately 55\,000 node hours for one time step. 

\subsection{Height stratification of atmospheric parameters}
In this section, the results from the STiC inversion are analyzed. The time step at 02min 55s was chosen, because the filament is changing shape around this time and the \ion{Ca}{II}~8542~\SI{}{\angstrom} blue wing brightening is strongest (see~Fig.~\ref{lightcurve}). 

The response functions (describing the sensitivity of the spectra at a certain wavelength point to the change of a given atmospheric parameter at a certain optical depth) are used to evaluate at which optical depth regions spectral information was available for the inversion code to determine atmospheric parameters. For temperature, the sensitivity was distributed roughly over the optical depth range between $\log\tau=0$ and $\log\tau=-5$. Outside this region, none of the lines used here showed substantial response values. For velocity and also for the magnetic field parameters, the region of sensitivity was roughly between $\log\tau=-0.5$ $\log\tau=-5.5$. We additionally emphasize that the result of our specifically adapted inversion strategy represent one possible solution that fit the complex profiles. We cannot exclude that the solution is degenerate, which means that there could be other solutions that fit the profiles. 

\subsubsection{Example profiles}
\begin{figure*}[h!]
\centering
	\includegraphics[width=18cm]{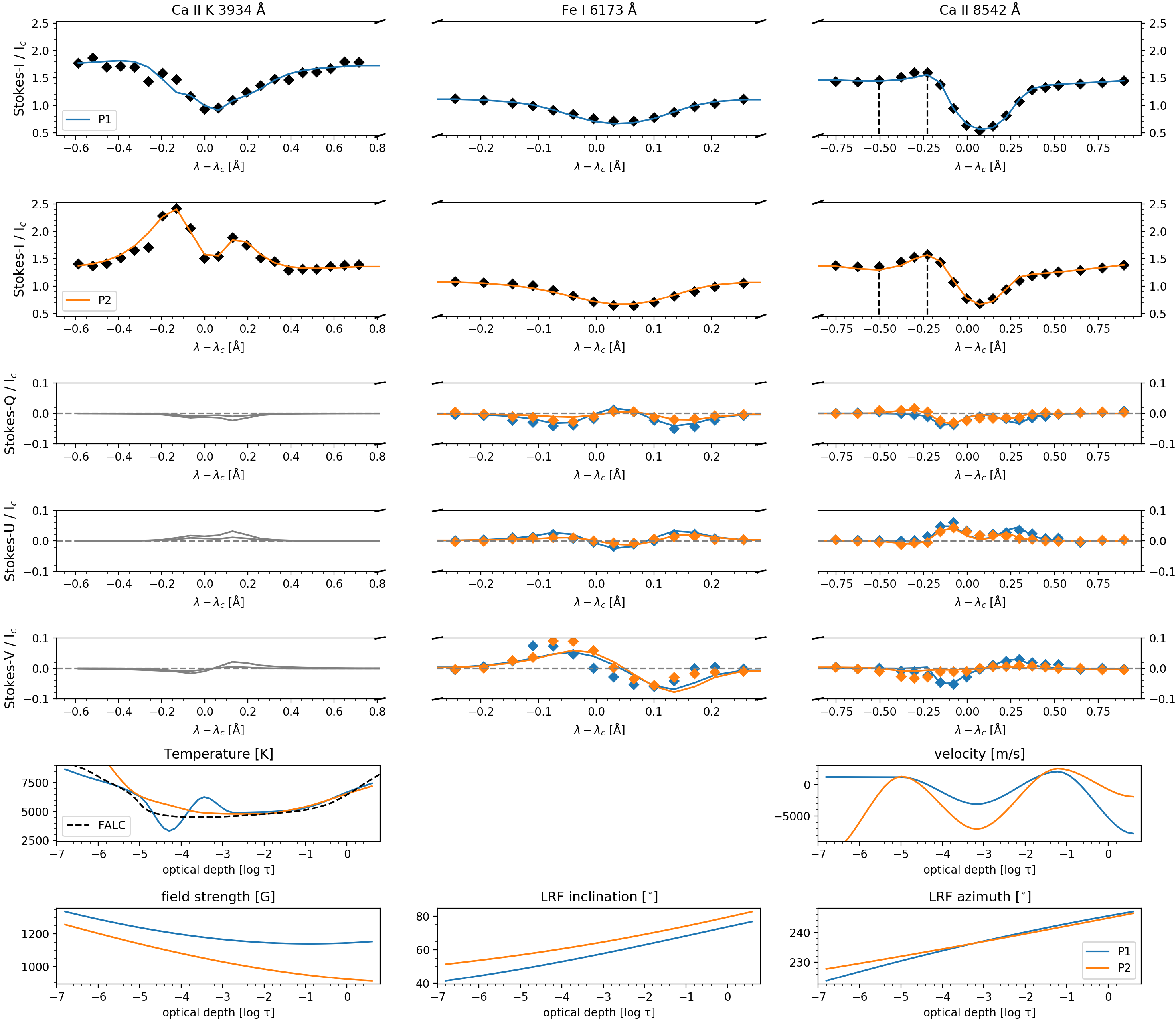}  
     \caption{Fit and corresponding depth-dependent atmospheric models from the STiC inversion for two example pixels P1 (blue solid lines) and P2 (orange solid lines). The observed wavelength points are shown as diamonds. For the \ion{Ca}{II}~K~3934~\SI{}{\angstrom} line, only the synthesized spectra are available in polarimetry (solid gray lines). The location of P1 and P2 is shown in Fig.~\ref{stic35}. The black dashed vertical lines in the \ion{Ca}{II}~8542~\SI{}{\angstrom} line line spectra depict the wavelength positions, from which the difference maps shown in Fig.~\ref{bspos} are calculated.} 
     \label{example}
\end{figure*}

Figure~\ref{example} shows both the fitted profiles and the corresponding atmospheric model resulting from the STiC inversion for two example pixels P1 and P2. For completeness, the stratification of the atmospheric parameters are also shown for optical depth values outside the sensitivity regions, but there, the atmospheric parameters can not be determined well by the inversion. 

Both pixels P1 and P2 show a BWP in the \ion{Ca}{II}~8542~\SI{}{\angstrom} line. Pixel P1 (plotted as blue lines) is an example of a complex pixel in which the \ion{Ca}{II}~K~3934~\SI{}{\angstrom} line is quiet (without clear emission peaks, cf. Sect.\,\ref{spectral_characterization}). Both spectra were successfully fitted by the code. The main difference between the atmospheric models is the temperature stratification. For reference, the temperature stratification from a FALC (quiet-sun) model atmosphere \citep{1993ApJ...406..319F} is shown. For P1, the code produced a temperature enhancement around $\log\tau=-3.5$, followed by a depression around $\log\tau=-4.5$. For P2, a temperature stratification with a plateau between $\log\tau=-2$ and $\log\tau=-3.5$ and no intermediate peaks was produced. Compared to the FALC atmosphere, P2 shows increased temperatures mostly in the height range between $\log\tau=-3.5$ and $\log\tau=-5$ (low chromosphere). At $\log\tau=-3.5$, the temperature difference between the FALC atmosphere and the inversion result was \SI{1760}{\kelvin} for P1 and \SI{380}{\kelvin} for P2. In layers above $\log\tau=-5$, the sensitivity is low and the stratifications deviate strongly. Below $\log\tau=-2.5$, the curves follow a similar stratification. 

The velocity stratifications are qualitatively similar in terms of both showing a central depression around $\log\tau=-3$ with negative values (flows approaching the observer). At around $\log\tau=-4.5$ and also around $\log\tau=-1.5$, positive values are reached. Above and below these optical depth values, the sensitivity decreases and the curves diverge. The main difference between the two pixels is that the amplitude of the main depression is larger for P2 than for P1.

\subsubsection{Two-dimensional maps in the high photosphere}
\begin{figure}[h!]
\centering
	\includegraphics[width=8.8cm]{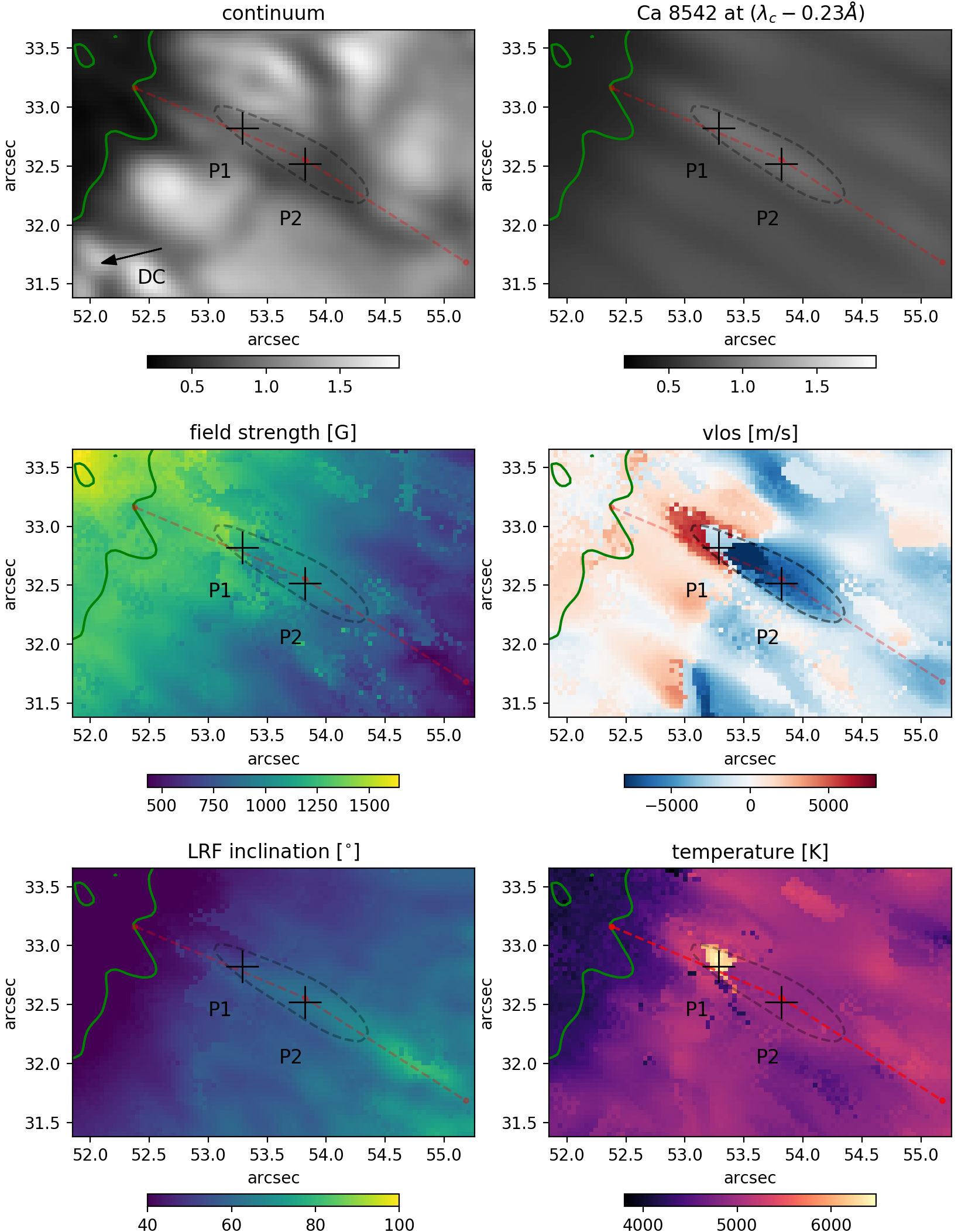}
     \caption{Intensity images and STiC atmospheric parameters at $\log\tau=-3.5$ from the time step at 02min 55sec. The gray dashed line represents a contour in the blue wing slope maps (see Fig.~\ref{bspos}) at the level of zero. The green contour outlines the pore boundary in continuum intensity. The value range of the inclination map was set to [40,100] to allow for direct comparisons to other maps. The black arrow in the bottom-left corner of the continuum maps points toward disk center. The green contour outlining the pore was produced from the CHROMIS  image at the continuum wavelength point at \SI{4000}{\angstrom} and is overplotted on the other panels.} 
     \label{stic35}
\end{figure}

In Fig.~\ref{stic35}, maps from the STiC results are shown at a constant optical depth value of $\log\tau=-3.5$ corresponding to the high photosphere. The location of the example pixels P1 and P2 are indicated with crosses. The gray dashed line in the LOS velocity (vlos) and temperature maps represents a contour from the blue-wing-slope map (see Fig.~\ref{bspos}) at zero level. Inside this contour, the values are positive, which means that a BWP is present. This is the case for almost the entire area of the \ion{Ca}{II}~8542~\SI{}{\angstrom} blue-wing brightening. In the velocity map, bidirectional flows are present in close vicinity. On the upper left part of the area enclosed by the contour, strong redshifts (up to \SI{8}{\kilo \meter \per \second} in velocity) are present. In a small region around the location of P1,  blueshifts with velocity values around \SI{3}{\kilo \meter \per \second} are seen while the remaining area enclosed by the contour includes strong blueshifts with values around \SI{8}{\kilo \meter \per \second}. At this heliocentric angle, blueshifts can represent upflows or also horizontal flows toward disk center (see black arrow in Fig.~\ref{stic35}). In temperature maps, a small region of increased temperature is located around P1 in correlation with the \SI{3}{\kilo \meter \per \second} blueshifts in the velocity map. We manually verified that in this region, the Stokes-$I$ spectra show the complex behavior of a BWP in the \ion{Ca}{II}~8542~\SI{}{\angstrom} line and a quiet \ion{Ca}{II}~K~3934~\SI{}{\angstrom} profile (cf. the spectra of P1 in Fig.~\ref{example}). The temperature difference between pixels in this region to pixels in the close surrounding was roughly \SI{1000}{Kelvin}. The magnetic field strength map, as expected, shows the largest values in the pore which decrease with distance from the pore boundary. In inclination, the filament is not as clearly visible at this atmospheric height as in the VFISV (low photosphere) inclination maps. Its right end, however, shows increased inclination values around \SI{90}{\degree}, as seen already in the VFISV inversion maps (Fig.~\ref{vfisv_fil}).

\subsubsection{Vertical cut through the atmosphere}
\label{analysis_cut}
\begin{figure}[h!]
\centering
	\includegraphics[width=8.8cm]{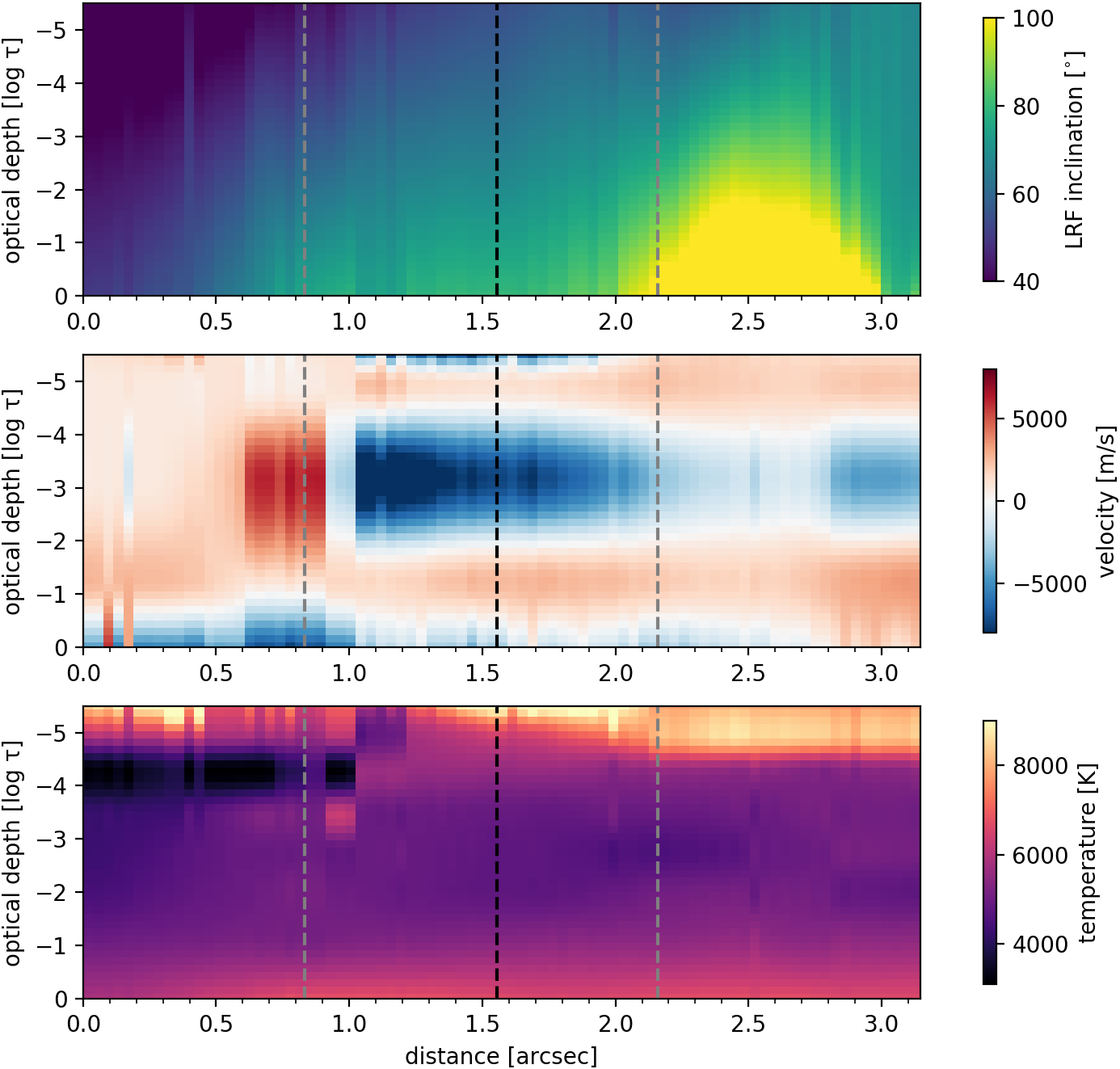}
     \caption{Vertical cut through the STiC model atmosphere along the red dashed line. The distance of \SI{0}{\arcsec} corresponds to the left end of the line (close to the pore) and the distance of \SI{3.1}{\arcsec} corresponds to its right end. The location of the kink in the red line (close to pixel P2) is depicted with vertical black dashed lines. The intersections of the gray contour from Fig.~\ref{stic35} with the red dashed line are depicted with gray dashed lines. The value range of the inclination map was set to [40,100] to allow for direct comparisons to other maps.} 
     \label{sticcut}
\end{figure}

In order to get an understanding of the vertical structure of the filament and the temperature increase in the chromosphere, a vertical cut through the model atmosphere from the STiC inversion is plotted in Fig.~\ref{sticcut}. The cut was produced along the red dashed line shown, e.g., in Fig.~\ref{stic35}. The distance of \SI{0}{\arcsec} corresponds to the left end of the line (close to the pore) and the distance of \SI{3.1}{\arcsec} corresponds to the right end. The limits of the displayed optical depth values was chosen to be $\log\tau=0$ and $\log\tau=-5.5$, which includes the sensitivity ranges of temperature, velocity and magnetic field inclination (see beginning of this subsection). 

The pattern of bidirectional flows already seen in Fig.~\ref{stic35} is clearly visible. At distances between \SI{0.6}{\arcsec} and \SI{0.9}{\arcsec}, redshifts are visible in the central region of the atmosphere between $\log\tau=-2$ and $\log\tau=-4$. This region corresponds mostly to the high photosphere. In this height range, a small region of blueshifts with velocity values around \SI{3}{\kilo \meter \per \second} is present at a distance of \SI{1}{\arcsec}. At distances beyond \SI{1}{\arcsec}, strong blueshifts are seen in the central part of the model atmosphere with values up to \SI{8}{\kilo \meter \per \second} and higher. With increasing distances, the values decrease. 

In the temperature cut, the region with weak blueshifts shows a temperature enhancement at $\log\tau=-3.5$, and a depression at $\log\tau=-4.5$. Pixel P1 shown in Fig.~\ref{example} is therefore representative of this region. Close to and inside the pore (below a distance of \SI{0.7}{\arcsec}), a clear temperature minimum is present around $\log\tau=-4.5$. This is not the case for distances beyond \SI{1.2}{\arcsec}. Until a distance of \SI{2}{\arcsec}, no steep vertical temperature gradients are visible in the plotted value range. Instead, there is a plateau with temperature values around \SI{5000}{\kelvin}. Pixel P2 shown in Fig.~\ref{example} shows such a temperature stratification which is characterized by an extended temperature increase in the low chromosphere, compared to a FALC reference atmosphere. For distances between \SI{2}{\arcsec} and \SI{2.5}{\arcsec}, increased temperatures around $\log\tau=-5.0$ are present, together with a more confined minimum around $\log\tau=-3.0$. In summary, enhanced temperatures are found in the high photosphere on the left side of the filament (close to the pore). With increasing distance from the pore, the temperature enhancement is located in higher layers (low chromosphere).

In  inclination, the lowest values are present on the left side (distance below \SI{0.5}{\arcsec}). This is followed by a region of increased inclination values throughout the atmosphere but with slightly higher values around \SI{70}{\degree} in the lower part of the atmosphere (below $\log\tau=-2$). Beyond a distance of \SI{2.0}{\arcsec}, a region with increased inclination values up to \SI{100}{\degree} is present in the lower atmosphere (below $\log\tau=-2$). A region with such values is also present in the VFISV inversions maps shown in Fig.~\ref{vfisv_fil}.

\section{Summary and Discussion}
We presented observations of a large solar pore and its immediate surroundings with relevant information of the physical properties of the evolution of transient photospheric filamentary features and their chromospheric counterparts.

\subsection{Structure and evolution of filaments in the photosphere}
In maps showing the inclination angle of the magnetic field, four radially aligned filaments with increased inclination values (up to \SI{70}{\degree} and more, see Fig.~\ref{vfisvmaps}) were seen. While blueshifts with values around \SI{2}{\kilo \meter \per \second} were present in the vicinity of all filaments, only one of those filaments (named F4) showed a conspicuous counterpart structure in continuum intensity images. 

At this position on the solar disk, blueshifts on the right-hand side of the pore can represent either upflows and/or horizontal radial inflows. A horizontal radially symmetric inflow pattern surrounding the pore \citep[as reported by e.g.][]{2010A&A...516A..91V} would be observed as blueshifts on the right-hand side and redshifts on the left-hand side of the pore. Although a weak tendency of such an inflow pattern could be recognized in Fig.~\ref{vfisvmaps}, the strong blueshifts with velocity values above \SI{1500}{\meter \per \second} are confined to the filaments only. We therefore assume that the blueshifts observed in the transient filaments represent upflows.

The finger-like structures observed by \citet[][ Fig.~3]{2008ApJ...676..698B} around a naked sunspot after the decay of a penumbra showed similar characteristics (blueshifts and inclined magnetic fields) to the filaments described here. They observed only `tiny fibrils' in G-band images, but no filamentary structures in continuum images. The authors speculated that the finger-like structures were located above the continuum forming layer and interpreted them as the remnants of decaying penumbral fields that are rising toward higher layers. The filaments observed here could also be similar to the `opposite polarity patches' around a pore with a decaying transient penumbra analyzed in magnetograms by \citet{2014ApJ...796...77W}. They ascribed those as remnant horizontal fields turning into vertical umbral field. However, in their case, only magnetograms and no vector magnetic field information nor velocities were given, so a direct comparison is difficult.

Filament F4 and its decay within less than \SI{10}{\minute} were analyzed in detail (see Fig~\ref{vfisv_fil}).  The evolution of the magnetic field vector showed that, at the beginning of the decay, the magnetic structure of the filament was similar to a magnetic $\Omega$-shaped loop (cf. the magnetic inclination configuration in Fig.~\ref{incalong}). We note, however, that the magnetic field at the outer footpoint of the loop is not vertical, but has inclination values of about \SI{110}{\degree}. 

The question arises whether the visible inclination profile (cf. Fig.~\ref{incalong}) can be produced by one flux tube loop, or only by several flux tubes aligned in parallel. In Fig.~\ref{incalong}, we consider the region between \SI{0.75}{\arcsec} and \SI{2.25}{\arcsec} as the region in between the footpoints. The average inclination in this region is \SI{73}{\degree}, which corresponds to a height gain of \SI{330}{\kilo \meter}. The vertical extent of such a flux tube would therefore be comparable to the height extent of the formation region of the \ion{Fe}{i}~6173~\SI{}{\angstrom} line \cite[at least \SI{300}{\kilo \meter}, according to][]{2011SoPh..271...27F}. It is therefore possible that the polarimetric signals are produced by one flux tube.

The loop-like feature associated with filament F4 dissolved during the decay. One way to explain the decay would be that the flux tube is rising upwards and dissolves there. This scenario would be in line with the observed blueshifts in the low photosphere. The newly forming structure seen in inclination and velocity maps at the last time step shown in Fig.~\ref{vfisv_fil} could be a sign of a new flux tube that is about to rise from below the surface.

\subsection{Reasons for missing penumbra}
In a larger context, the question arises why the transients filaments have not transformed into penumbral filaments, that is, why the observed pore has not developed into a sunspot with a stable penumbra. This question is also motivated by the above-mentioned similarity of the present transient filaments to the finger-like features observed by \citet{2008ApJ...676..698B}. The finger-like structures have been observed after the decay of a penumbra and could, as suggested by the authors, represent remnant penumbral field lines that rise upwards. For sunspots with stable penumbrae, there are hints toward a chromospheric canopy that confines (`traps') magnetic field lines in the photosphere preventing them from further rising (\citet{1998ApJ...507..454L}; \citet{2013ApJ...769L..18L}; \citet{2018ApJ...857...21L}; \citet{2019ApJ...886..149L}; and Lindner et al. (2023, [proper reference pending])). In the case of the observed pore, however, no signs of a large-scale chromospheric canopy are identified (see, e.g. Fig.~\ref{stic35}). This could lead to the flux tube of the transient filaments being able to freely rise further toward higher layers. A missing chromospheric canopy could therefore be a reason for the decay of the transient filaments and therefore for the missing penumbra. 

As mentioned in Sect.~\ref{hmi}, the magnetic flux value of the pore is in the transition range between that of a pore and that of a sunspot. The limited magnetic flux of the pore is therefore a second possible reason for a missing stable penumbra. 

In general, the two requirements of enough magnetic flux and a chromospheric canopy could also be connected: If the main flux tube of a pore/protospot gets larger, the field lines in chromospheric layers might fan out more. This fanning out of field lines could then supply the horizontal magnetic fields that a chromospheric canopy is made of.

\subsection{Signs of magnetic reconnection in high photosphere and low chromosphere}
At the location of the filament F4 (shifted by approximately \SI{0.15}{\arcsec} toward north), an elongated brightening in the blue wing of the \ion{Ca}{II}~8542~\SI{}{\angstrom} line was observed. This brightening was shown to be caused by a peak in the blue line wing (see Fig.~\ref{spec_along}), abbreviated as BWP. This spectral feature appeared co-temporal and co-spatial to the decaying filament. Intensity difference maps at different wavelength positions (Fig.~\ref{bspos}) additionally showed that such BWP profiles are also present in the vicinity of other filaments. Therefore, a causal connection between the decaying filament F4 and this brightening is likely. BWPs have also been shown to be a common spectral feature in penumbral microjets \citep[see, e.g.][Fig.~2]{2019ApJ...870...88E} and other reconnection events \citep[e.g.][Fig.~5]{2021A&A...647A.188D}, sometimes accompanied by peaks in the red line wing. Profiles with similar emission peaks (only on the red wing) have also been observed in emerging-flux regions \citep{2015ApJ...810..145D}. 

A specific inversion strategy using the STiC inversion code was used to obtain atmospheric parameters over a broad atmospheric height range from three spectral lines. The region around filament F4 was inverted with STiC for one time step. The maps of the LRF inclination of the magnetic field are consistent with the corresponding map from the VFISV inversion that used the \ion{Fe}{I}~6173~\SI{}{\angstrom} line only: A filament with increased inclination values and maximum values at its outer end can be seen in Fig.~\ref{stic35} and Fig.~\ref{sticcut}. 

In the temperature maps at $\log\tau=-3.5$, a confined region with increased temperature is located at the left side of the \ion{Ca}{II}~8542~\SI{}{\angstrom} brightening (around pixel P1). The difference with respect to the close surrounding is about \SI{1000}{\kelvin}. This region is located exactly in between a region with strong redshifts and a region with strong blueshifts, both with amplitudes up to \SI{8}{\kilo \meter \per \second}. The vertical cut through the atmosphere (Fig.~\ref{sticcut}) shows that the bidirectional flows are present approximately between $\log\tau=-2$ and $\log\tau=-4$. The region in between the strong up- and downflows is characterized by weak blueshifts and the aforementioned temperature increase around $\log\tau=-3.5$ and a temperature minimum around $\log\tau=-4.5$. Pixel P1 shown in Fig.~\ref{example} shows an example of such a pixel. We note that the amplitude and width of the temperature enhancement and depression could be overestimated by the inversion, that only has a limited number of nodes available. An even more dense location of nodes might have resulted in a smaller and more localized peak and no depression. We also note that the red dashed line, along which the vertical cut was produced, only cuts trough a few pixels of the temperature increase region (see Fig.~\ref{stic35}). If the line had been shifted upwards by a few pixels, the respective temperature increase region in the cut would have been larger. We kept the location of the red line to be able to compare to the photospheric results from the VFISV inversions, that were produced along the same line (see Fig.~\ref{incalong}). 

Bidirectional flows, often together with temperature increases, have previously been reported in connection with observations of, among others, Ellermann bombs and ultraviolet bursts \citep[e.g.][]{2019A&A...627A.101V}, microflares \citep[e.g.][]{2016ApJ...820L..17H}, flares and X-Ray bursts \citep[e.g.][]{2020ApJ...900...17Y}. \citet{2022FrASS...920183S} gives an overview over some of those events. They are mostly interpreted as signs of magnetic reconnection. The location of the reconnection event is sometimes difficult to  determine and is believed to be situated, depending on the event, at different atmospheric heights from the high photosphere to the corona. 

The temperature and velocity signals shown in Fig.~\ref{stic35} and Fig.~\ref{sticcut} are partially similar to the ones reported by \citet{2021A&A...647A.188D} which were also observed close to a pore. In their case, the magnetic reconnection event was believed to be triggered by magnetic flux cancellation occurring in an extended region with an area of \SI{6}{\arcsec}x\,\SI{6}{\arcsec}. In our case, the signs of reconnection are visible on a smaller scale. The temporal evolution of the decaying filament and the \ion{Ca}{II}~8542~\SI{}{\angstrom} blue wing brightening (see Fig.~\ref{vfisv_fil}) suggest that the reconnection event is connected to the decay of the single filament F4. Recently, \citet{2022ApJ...925...46S} showed imaging data of a surge in H-$\alpha$ that developed after the decay of an abnormal granule in the photosphere. Due to the lack of spectroscopic data, however, no velocity information could be analyzed. We are not aware of a publication in which spectropolarimetric data of a reconnection event has been analyzed in connection with such a small scale single (almost isolated) photospheric feature. 

\subsection{Model describing the magnetic reconnection}
\label{models_recon}
Our observations show signs of magnetic reconnection in connection to a single almost-horizontal flux tube, which provides a good example to directly compare to simulations. The most common models describing the reconnection between new emerging and pre-existing fields are based on the model by \citet{1995Natur.375...42Y} (developed following the idea of \citet{1977ApJ...216..123H}), which includes an `anemone' shaped magnetic configuration for events happening in active solar regions. This model describes reconnection events that happen on spatial scales of several tens of \SI{}{\mega \meter} centered in the corona. More recent simulations have been performed, e.g., by \citet{2016ApJ...822...18N}. \citet{2021A&A...647A.188D} proposed that a small `anemone' type reconnection event can also happen in the low chromosphere to explain their observed flows and temperature enhancement. In our case, the \ion{Ca}{II}~8542~\SI{}{\angstrom} brightening, the inferred bidirectional flows and the temperature enhancement occurs on an even smaller scale. The magnetic field of filament F4 shows a region with opposite polarity and this region is located at the outer end of the filament. The `anemone' model proposed by \citet[][Fig. 11]{2021A&A...647A.188D}, however, includes a magnetic loop with the opposite polarity at the edge of a pore, where the magnetic reconnection happens. Therefore, we propose a different mechanism for the magnetic reconnection happening in our case. 

A possible scenario would be that a magnetic flux tube is moving from the deep photosphere toward the high photosphere, possibly because no chromospheric canopy stops the rise. This is supported by the blueshifts observed in the photosphere (Fig.~\ref{vfisv_fil}). The gas pressure decreases and the flux tube expands. This leads to an instability of the flux tube and at some point, magnetic reconnection can happen at its edges, where the magnetic field is sheared with respect to the background magnetic field. This would be similar to reconnection events at the boundary of single penumbral filaments that have been observed as penumbral microjets, e.g. by \citet{2019ApJ...870...88E}. In our case, the reconnection of field lines would change the magnetic field configurations at the edge of the flux tube, which accelerates the instability of the flux tube and the filament decays further. The temperature increase observed in the STiC results (e.g. Fig.~\ref{stic35}) can be explained by the transformation of magnetic energy into thermal energy that happens during the reconnection. Explaining the bidirectional flows (downflows on the left side of the filament, upflows on the right side and a region with reduced velocities in between) is more complex. 

\begin{figure}[h!]
\centering
	\includegraphics[width=8.8cm]{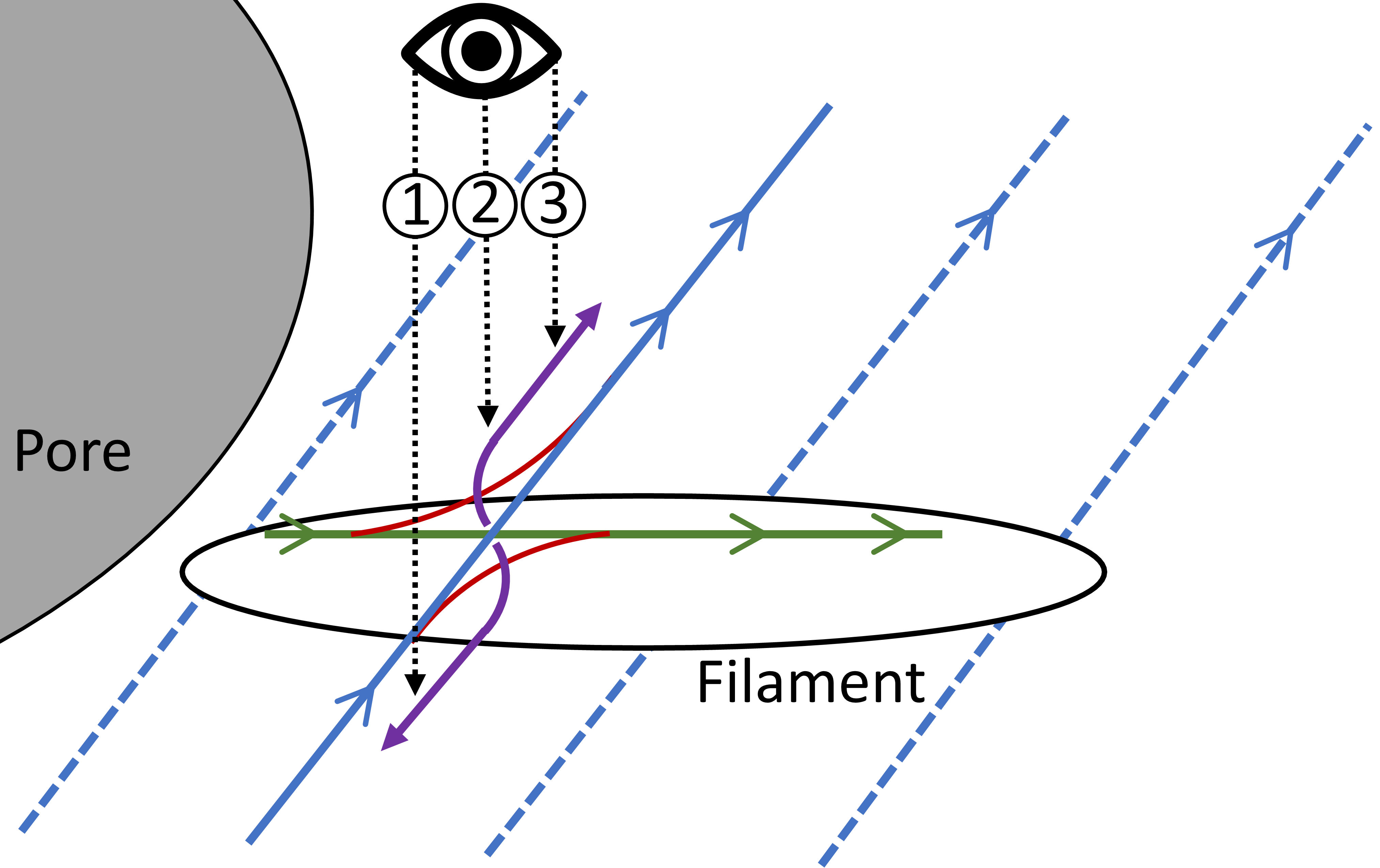}
     \caption{Sketch of the proposed reconnection event. Blue lines: Weakly inclined background field magnetic field. Green line: Mostly horizontal field of filament. Red lines: Newly connecting field lines forming during the reconnection event. Arrows indicate the direction of the magnetic field lines. Purple curved arrows: Path of the plasma that initially gets accelerated due to magnetic tension and then gets deflected by the surrounding background field lines.} 
     \label{recon}
\end{figure}

We propose four mechanisms on how magnetic reconnection can lead to the observed flow field. It is also possible that several of these mechanisms work in combination. A sketch of the relevant magnetic fields is shown in Fig.~\ref{recon}. The background magnetic field lines are shown as blue lines and the magnetic field lines of the flux tube as green lines. The new field line connections are shown as red lines. Black dashed lines show the viewing paths of an observation. We neglect projection effects, because the angle between the LOS and the axis of the filament is around \SI{45}{\degree}. 
We note that the enhancement and the following depression in the temperature stratification of Pixel P1 (see Fig.~\ref{example}) could, in principle, also be caused by an acoustic shock similar to the model introduced by \citet{1997ApJ...481..500C} and recently observed by \citet{2022A&A...668A.153M}. However, we limit the discussion to magnetic reconnection as the driver of the observed temperature and velocity signals because these two observed signatures occur co-spatial and co-temporal to the decay of the magnetic transient filament and we also do not identify a mechanism how the shock front would also explain the observed flow field.

\begin{enumerate}
\item Similar to a rubber band, there is a magnetic tension in the new field line connections (red lines in Fig.~\ref{recon}). The force is directed orthogonal to the red lines and leads to an acceleration of the neighboring plasma. The accelerated plasma, however, immediately encounters the background field and gets deflected to move along the background field lines. These flow paths are shown as purple arrows in Fig.~\ref{recon}. On the left side of the reconnection event (viewing path 1), downflows would be observed. In the center of the reconnection event (viewing path 2), both up and downflows are present and the resulting combination depends on the exact orientation of the relevant magnetic field lines and the exact viewing angle. In the case of filament F4, a region with increased high photospheric temperatures and weak blueshifts are observed between the bidirectional flows. On the right side of the reconnection event (viewing path 3), upflows would be observed. 
\item The sudden temperature increase in the volume around the reconnection event leads to a pressure gradient with respect to the surrounding plasma. The heated plasma moves away from the reconnection site to balance this pressure gradient. At the reconnection site it has the two options of either moving along the weakly inclined background magnetic field or along the inclined magnetic field of the filament. Since the background field is present in a larger volume than the field of the spatially confined filament, the motion along the background field dominates. The upflows and downflows would, again, be consistent with the observations.
\item The newly connecting field lines lead to the formation of two new flux tubes. Following the direction of the magnetic field, one starts at the bottom with the former background field (blue), then continues along the new field line (red) and ends with the field line of the former filament (green). According to the colors of the field lines shown in Fig.~\ref{recon}, we call this new flux tube `BRG'. Similarly, there is a newly forming flux tube `GRB'. We assume that, before the reconnection event, the system was in hydrostatic equilibrium and the magnetic field lines in pressure equilibrium, that is, $p_g + \frac{B^2}{2\mu_0}=\mathrm{const}$ with $\mu_0$ being the electromagnetic permeability. Since the magnetic field strength $B$ of the background field is assumed to be larger than the field of the filament, the gas pressure $p_g$ must be smaller. This leads to a pressure gradient inside the new flux tubes. For BRG, this leads to force which is anti-parallel to the magnetic field direction and a respective flow along BRG. For GRB, the force is parallel to the magnetic field direction and a respective flow along GRB develops. In an idealized situation, an observer with a viewing angle from above would not measure the horizontal flows, but only the flows with vertical components. Again, downflows on the left of the filament would be observed (start of `BRG') and upflows on the right (end of `GRB'). At the reconnection site, a combination would be observed.
\item The filament is observed to be a bright structure and is therefore assumed to harbor plasma with higher temperature than in the surrounding. Inside the new flux tubes `BRG' and `GRB', there is now an end with higher temperatures and an end with lower temperatures. Again, a pressure gradient develops leading to the same flow topology as for mechanism 3.
\end{enumerate}

Mechanism 1 and 2 would imply flows of heated plasma parallel to the background field components. This could explain why temperature enhancements were observed at increasing height with increasing distance from the reconnection site (see Sect.~\ref{analysis_cut}).

We suggest that the above mentioned reconnection process is not a single event. Instead, it is taking place at the boundary surface between the flux tube of the filament and the surrounding over a certain time interval on the order of minutes. This would explain why the brightening observed in the blue wing of the \ion{Ca}{II}~8542~\SI{}{\angstrom} line was present already in the beginning of the time series and lasted for about \SI{5}{\minute} (see Fig.~\ref{lightcurve}).

The differences to the reconnection events ascribed to penumbral microjets would be that penumbral filaments in general remain stable, while the filament described here is rising toward the high photosphere. Therefore, it is more vulnerable to magnetic reconnection at its edges than penumbral filaments located in the deep photosphere, where the gas pressure is higher than the magnetic pressure. In addition, the transient filament is not embedded in a penumbral structure, but is standing more isolated. The magnetic field at its edge can therefore be more sheared than the magnetic field between spines and intraspines of the penumbra, allowing for stronger reconnection events. This would explain why observed velocities in penumbral microjets are lower than the velocities reported here \citep{2019ApJ...870...88E}.

Magnetic reconnection at the edge of the filament would also explain why the temperature increase seen in the upper photosphere (Fig.~\ref{stic35}) is located at the northern edge of the filament and not at its center. An alternative explanation for the shift of the Ca IR brightening and the temperature increase region with respect to the photospheric filament axis would be that reconnection in the low or mid photosphere still happens in the center of the filament. Starting from there, the hot plasma is transported into the chromosphere along the closest pre-existing fibril. This fibril does not need to be located directly above the filament axis, but can be slightly shifted. Inside this fibril, the temperature increases and the up and downflows are aligned along the fibril. In addition, projection effects due to the heliocentric angle could contribute to small observed shifts in the image plane between structures at different geometric heights.

\section{Conclusions}
We have presented a detailed analysis of the decay of a transient filament at the outskirts of a large pore and its chromospheric counterpart observed as an emission peak in the blue wing of the \ion{Ca}{II}~8542~\SI{}{\angstrom} line. From the temperature increase (about \SI{1000}{\kelvin}) and the bidirectional flows (up to \SI{8}{\kilo \meter \per \second} in magnitude) in the high photospheric/low chromospheric inferred from the multi-line inversions with the STiC code, we identify magnetic reconnection as the most likely mechanism behind the decay of the transient filament and present different detailed scenarios explaining the observations.

In addition to the above mentioned transient filament, other such transient filaments around the pore are observed. They show a similar chromospheric response indicating that they possibly also undergo magnetic reconnection. Yet, the spectral signature of those other features is weaker and their analysis would require a different, specifically adapted inversion strategy. Studying more such cases would be necessary to find out how frequent such events are. 
On the one hand, the investigation of such transient (single) filaments might be key to understanding the necessary conditions for the formation and decay of penumbral filaments in stable sunspots. On the other hand, if such reconnection events happen regularly in the vicinity of pores and sunspots, the resulting ejection of hot plasma into the chromosphere can be considered as one of the mechanisms contributing to the ongoing heating of the chromosphere.

\begin{acknowledgements} 
We thank Oleksii Andriienko for his help during the observations and also for calibrating the SST data and providing the data to us. We also thank Mats Löfdahl for his help in solving problems with the data reduction. We thank Vigeesh Gangadharan for his useful feedback on the manuscript. JdlCR gratefully acknowledges funding from the European Research Council (ERC) under the European Union's Horizon 2020 research and innovation program (SUNMAG, grant agreement 759548).
The Swedish 1-m Solar Telescope is operated on the island of La Palma by the Institute for Solar Physics of Stockholm University in the Spanish Observatorio del Roque de los Muchachos of the Instituto de Astrofísica de Canarias. The Institute for Solar Physics is supported by a grant for research infrastructures of national importance from the Swedish Research Council (registration number 2021-00169). \\
This project has received funding from the European Union's Horizon 2020 research and innovation programme under grant agreement No 824135, the Trans-National Access Programme of SOLARNET. 
\end{acknowledgements}

\begin{spacing}{0.81}
    { \let\par\relax
    \noindent 
    \tiny 
    \textit{Author contribution statement.} 
    PL designed and directed the research project with support from NBG and RS. 
    PL conducted the SST observations together with NBG. 
    JdlCR processed the SST data together with his team and with support from PL. 
    PL ran the VFISV inversions. 
    PL analyzed the transient filaments, the BWP brightening, and the respective spectra with support from NBG and RS. 
    PL ran the STiC inversions together with JdlCR. 
    PL analyzed the STiC inversions with support from NBG, RS and JdlCR. 
    PL composed the proposed models describing the magnetic reconnection together with RS. 
    PL wrote the manuscript with support from RS, NBG and JdlCR. 
    }
\end{spacing}

\bibliographystyle{bibtex/aa} 
\bibliography{transfilambib} 

\end{document}